\documentclass[12pt]{article}
\usepackage{amscd,amssymb,amsmath,latexsym,enumerate}
\usepackage[mathscr]{euscript}
\usepackage{mathrsfs}
\usepackage{epsfig}
\usepackage{fancybox}
\usepackage{verbatim}
\usepackage{tikz}
\usepackage{tikz-cd}
\usepackage{todonotes}
\usepackage{multicol}
\usepackage{graphicx}
\usepackage{mathtools}

\usepackage{color}

\usepackage{xcolor}
\definecolor{MyBlue}{cmyk}{1,0.13,0,0.63}
\definecolor{MyGreen}{cmyk}{0.91,0,0.88,0.52}
\newcommand{\mylinkcolor}{MyBlue}
\newcommand{\mycitecolor}{MyGreen}
\newcommand{\myurlcolor}{black}

\usepackage{hyperref}
\hypersetup{%
  bookmarksnumbered=true,bookmarksopen=false,%
  plainpages=false,
  linktocpage=true,%
  colorlinks=true,breaklinks=true,%
  linkcolor=\mylinkcolor,citecolor=\mycitecolor,urlcolor=\myurlcolor,%
  pdfpagelayout=OneColumn,%
  pageanchor=true,%
}

\title{
Local perturbations of block Toeplitz matrices  
}

\author{Lars Koekenbier, Hermann Schulz-Baldes
\\
\\
{\small Friedrich-Alexander-Universit\"at Erlangen-N\"urnberg}
\\
{\small Department Mathematik, Cauerstr.~11, D-91058 Erlangen, Germany}
}

\date{ }

\newtheorem{theorem}{Theorem}
\newtheorem{proposition}[theorem]{Proposition}
\newtheorem{lemma}[theorem]{Lemma}

\newtheorem{remark}[theorem]{Remark}

\newcommand{\CM}{{\mathbb C}}

\newcommand{\SM}{{\mathbb S}}

\newcommand{\Aa}{{\cal A}}
\newcommand{\Ee}{{\cal E}}

\newcommand{\Ff}{{\cal F}}

\newcommand{\Ss}{{\cal S}}
\newcommand{\Oo}{{\cal O}}
\newcommand{\Tt}{{\cal T}}
\newcommand{\Rr}{{\cal R}}

\newcommand{\Mm}{{\cal M}}

\newcommand{\one}{{\bf 1}}

\newcommand{\spec}{\mbox{\rm spec}}

\newcommand{\Ker}{{\rm Ker}} 
\newcommand{\Ran}{{\rm Ran}}

\newcommand{\diag}{{\rm diag}} 
 
\newcommand{\Wind}{{\rm Wind}} 
\newcommand{\rk}{{\rm rk}}

\newcommand{\Toep}{{\mbox{\rm Toep}}}
\newcommand{\DToep}{{\mbox{\rm DToep}}}

\newcommand{\IndSet}{I}

\newcommand{\LimitSet}{\Lambda}

\begin{document}

\maketitle

\begin{abstract}
This work is about the asymptotic spectral theory of tridiagonal Toeplitz matrices with matrix entries, with periodicity broken on a finite number of entries. Varying the ranks of these perturbations allow to interpolate between open boundary and circulant Toeplitz matrices.  While the continuous part of the limit spectrum only depends on  these ranks and no other aspect of the perturbation, the outliers of the spectrum depend continuously on the local perturbation. The proof is essentially based on a new generalized Widom formula for the characteristic polynomial. All this holds for Lebesgue almost all perturbed Toeplitz matrices, a fact that constitutes another important extension of Widom's work. The mathematical results are illustrated by numerics.
\hfill 





\end{abstract}




\section{Overview}
\label{sec-Intro}

This paper shows how to compute the limit spectra of non-hermitian block tridiagonal operators of the form 
\begin{align}
\label{eq-ToeplitzIntro}
H_N(A,B,C) 
\, = & \; 
\begin{pmatrix}
C & T & & & & A \\
R & V & & & & \\
 & & \ddots & \ddots & & \\
 & & \ddots & \ddots & & \\
 &  & & & & T \\
B &  & & & R & V \\
\end{pmatrix}
\;.
\end{align} 
Here the matrix entries $R$, $T$ and $V$ as well as $A$, $B$ and $C$ are all $L\times L$ matrices with complex entries, and $H_N(A,B,C)$ is an $N \times N$ matrix with such matrix entries so that it is a $NL\times NL$ matrix with complex entries. Throughout, the coefficient matrices $R$, $T$ and $V$ will be kept fixed and it will be assumed that $R$ and $T$ are invertible. The matrices $A$, $B$ and $C$ specify a local perturbation and the object of this work is to study the dependence of its spectrum on this perturbation in the limit $N\to\infty$. Several interesting special cases will be considered: 

\begin{itemize}
\item[{\rm (a)}] 
$H_N(R,T,V)$ is the circulant (or periodic) Toeplitz matrix.

\item[{\rm (b)}]  $H_N(0,0,V)$ is the Toeplitz matrix (truncated, compressed, or open boundary conditions).

\item[{\rm (c)}] $H_N(0,0,C)$  is a boundary perturbation of the Toeplitz matrix from (b).

\item[{\rm (d)}] $H_N(A,B,C)$ with $B$ invertible will be referred to as a (semi-circulant or semi-permeable) local perturbation of the Toeplitz matrix. 

\end{itemize}

{In all cases, this work provides explicit formulas for an appropriate limit of the sequence of spectra $\Ss_N= \spec\big(H_N(A,B,C)\big)$. Recall \cite{BS,Bee} that for any sequence $(\Ss_N)_{N\geq 1}$ of closed subsets $\Ss_N\subset \CM$ one defines the inferior limit sets (also called the lower closed limits \cite{Bee} or the uniform limits \cite{BS}) and superior limit sets (also  called the upper closed limits \cite{Bee} or partial limits \cite{BS}) by
\begin{align}
\label{eq-InfLimit}
&
\liminf_{N\to\infty} \Ss_N 
\;=\; 
\big\{E\in\CM\,:\,
\exists\; E_N\in\Ss_N\;\mbox{\rm with }\lim_{N\to\infty}E_N=E
\big\}
\;,
\\
&\label{eq-SupLimit}
\limsup_{N\to\infty} \Ss_N 
\;=\; 
\big\{E\in\CM\,:\,
\exists\; N_k>N_{k-1} \;\mbox{\rm with } E_{N_k}\in\Ss_{N_k}\;\mbox{\rm with }\lim_{k\to\infty}E_{N_k}=E
\big\}
\;.
\end{align}
Both limit sets are known to be closed \cite[Section 5.2]{Bee}. Whenever they coincide,  then the sequence is said to converge in the Painlev\'e-Kuratowski sense and the limit set will simply be denoted by $\lim_{N\to\infty} \Ss_N$. In the main results below, this type of convergence will be proved and therefore the statements only involve $\lim_{N\to\infty} \spec\big(H_N(A,B,C)\big)$. Moreover, as $\spec\big(H_N(A,B,C)\big)$ is a uniformly bounded set, it is then also known that the convergence holds in the Hausdorff distance (see \cite{Bee}).
}

\vspace{.2cm}

\begin{figure}[h]
\centering
\includegraphics[width=5.5cm]{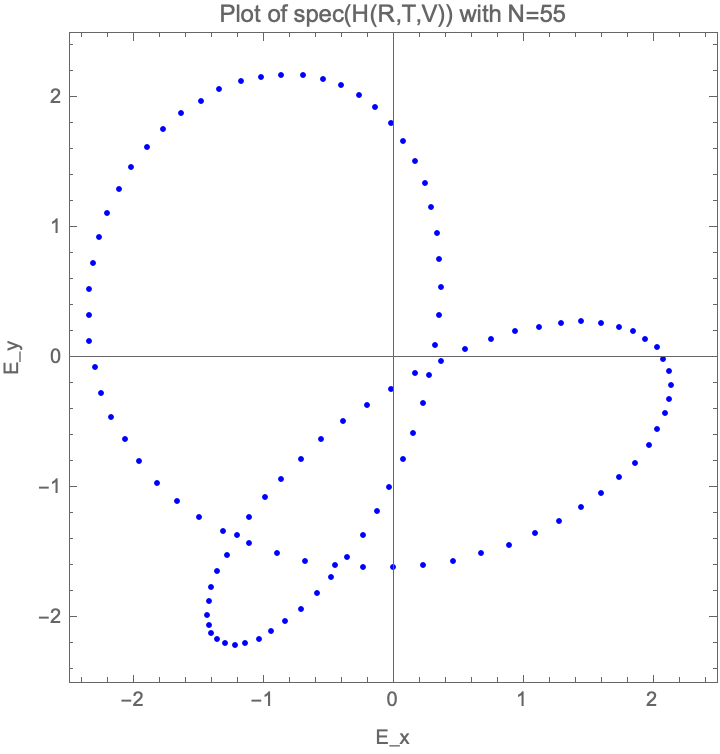}
\includegraphics[width=5.5cm]{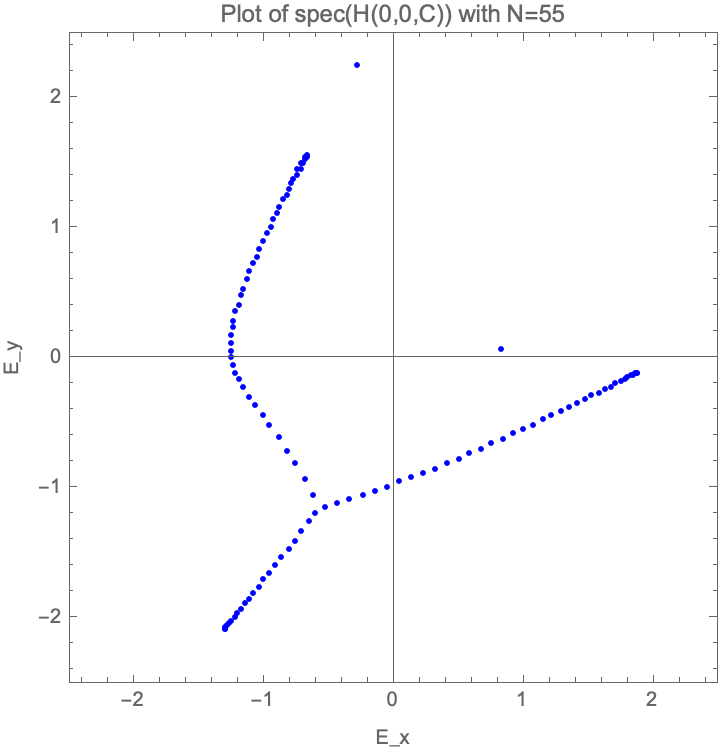}
\includegraphics[width=5.5cm]{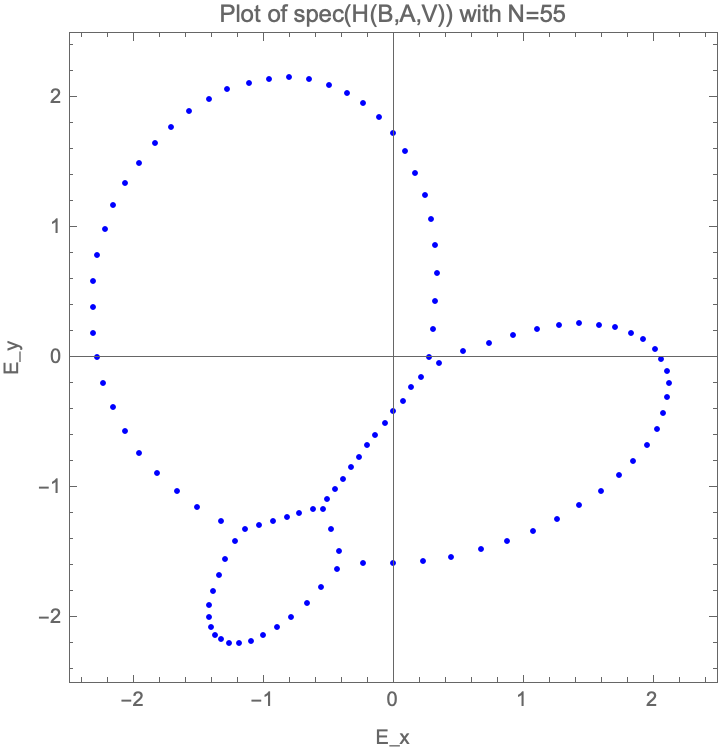}
\caption{\sl 
Numerical plot of the spectrum of $H_N(R,T,V)$, $H_N(0,0,C)$ and $H_N(B,A,V)$ for $N=55$, with all $2\times 2$ matrices $R,T,V,A,B,C$ chosen as in {\rm Section~\ref{sec-Numerics}}.
}
\label{fig-0}
\end{figure}

It is a well-known classical fact that the spectrum in  case (a) can readily be computed via the discrete Fourier transform \cite{BG,Nik}, which for a generic example with $L=2$ is shown in the first plot in Fig.~\ref{fig-0}. Even though all other cases are merely finite rank perturbations, the spectra  both for finite $N$ and in the asymptotic limit $N\to\infty$ look drastically different. Case (b) was studied by Widom \cite{Wid1} and more recently revisited in \cite{Del,KS} (the scalar case $L=1$ {goes back to Schmidt and Spitzer \cite{SS} and it} is extensively studied, see \cite{BG} and references therein). Under suitable conditions, these works provide an explicit formula for the limit spectrum, which generically consists of a tree-like structure and some outliers like in the second plot in Fig.~\ref{fig-0}. Then case (c) is a diagonal perturbation and it is shown in Theorem~\ref{theo-boundary} that its limit spectra is the same as in case (b) with modified outliers (note the two isolated eigenvalues in the plot). While Theorem~\ref{theo-boundary} is new, it follows from a minor modification of the techniques of \cite{KS}. 

\vspace{.2cm}

{Case (d) is considered in Theorem~\ref{thm-LimitSpecIntro}, which is the main new finding of this work. The entries $(A,B,C)$ allow to interpolate between case (a), (b) and (c).  Numerous works have studied the effect of such perturbations in the scalar case $L=1$, {\it e.g.} \cite{BW,BFGM,DK} and references therein, but we are not aware of  results on the matrix-valued case. Theorem~\ref{thm-LimitSpecIntro} provides an explicit expression for the limit spectra in case (d) and shows that its continuous part only depends on the rank of $A$. A numerical example is provided in the third plot of Fig.~\ref{fig-0}. One sees that part of the limit spectrum coincides with spectrum of the circulant Toeplitz matrix, while other parts converge towards a new limit curve (namely, the enclosed part attached by three $Y$-junctions).  Further numerical illustrations are given in  Section~\ref{sec-Numerics}.}

\vspace{.2cm}

\vspace{.2cm}

As in the prior work \cite{KS}, the central object of the analysis are the $2L\times 2L$ transfer matrices at complex energy $E\in\CM$ defined by
\begin{equation}
\label{eq-TraDef}
\Tt^E
\;=\;
\begin{pmatrix}
(E\,\one-V)T^{-1} & - R \\ T^{-1} & 0 
\end{pmatrix}
\;,
\end{equation}
which allow to compute the solutions of the eigenvalue equation locally, see the proof of Propositon~\ref{prop-CharPol}. The results below state that the continuous part of the limit spectrum can be computed from $\Tt^E$, albeit in a highly non-trivial manner, and that merely the discrete part of the limit spectrum, also referred to as outliers, depend on the local perturbation $(A,B,C)$. For example, the limit spectrum of the circulant Toeplitz matrices, case (a), are given by
\begin{equation}
\label{eq-PeriodicSpec1}
\lim_{N\to\infty} \spec(H_N(R,V,T))
\;=\;
\Sigma\;,
\end{equation}
where $\Sigma$ is given in terms of the spectral theory of the transfer matrices by
\begin{equation}
\label{eq-PeriodicSpec2}
\Sigma
\;=\;
\big\{ E\in\CM\,:\,\spec(\Tt^E)\cap \SM^1\not= \emptyset\big\}
\;.
\end{equation}
In text books like \cite{BS} one rather finds $\Sigma=\{ \spec(H(z))\,:\,z\in\SM^1\}$ where $H(z)$ is the symbol defined by $z\in\CM\mapsto H(z)=R z^{-1}+V+Tz$, but it is straight-forward to check that this coincides with \eqref{eq-PeriodicSpec2}. Determining the limit spectrum in the case (b) is Widom's achievement \cite{Wid1} which was rederived by transfer matrix techniques in the previous work \cite{KS}. It is the special situation $C=V$ of case (c) which is described in detail next. Some notations have to be introduced. Let $z_1(E),\ldots,z_{2L}(E)$ be the eigenvalues of $\Tt^E$, ordered such that $|z_i(E)|\leq |z_{i+1}(E)|$ for $i=1,\ldots,2L-1$. Degeneracies are collected in the set
\begin{equation}
\label{eq-FDef}
\Ff
\;=\;
\big\{E\in\CM\,:\,{\Tt}^E\mbox{ has a degenerate eigenvalue}\big\}
\;.
\end{equation}
As in \cite{KS}, it will always be assumed that the set $\mathcal{F}$ is finite.  Given a subset $I\subset\{1,\ldots,2L\}$, let $\Rr^E_I$  denote the Riesz projection of $\Tt^E$ onto the eigenvalues $z_j(E)$ with $j\in I$: 
\begin{equation}
\label{eq-RI}
{\Rr}_\IndSet^E
\;=\;
\oint_{{\gamma}_I}\frac{dz}{2\pi\imath}\;(z\,\one-\Tt^E)^{-1}
\;,
\end{equation}
where $\gamma_I$ is a curve with winding number $1$ around each  $z_j(E)$ with $j\in I$, but winding $0$ around  $z_j(E)$ with $j\not\in I$. For the finite number of $E\in\Ff$ and more generally in case there are eigenvalues $z_j(E)$ of the same modulus, there are some ambiguities resulting from the choice of eigenvalue labelling with the set $I$, but these ambiguities turn out to be irrelevant in what follows. Introduce the functions
\begin{equation}
\label{eq-qDef}
\hat{q}_{I}(E)
\;=\;
{\det}_L\Big(
\binom{0}{\one}^*
{\Rr}_{I}^E
\begin{pmatrix}
E\,\one-C & - \one \\ \one & 0 
\end{pmatrix}
\binom{\one}{0}\Big)
\;,
\qquad
E\in\CM\setminus\Ff
\;.
\end{equation}
The next theorem, corresponding to the asymptotic spectral analysis of case (c) in the above list, generalizes the results of \cite{Wid1,KS} and shows that a local diagonal contribution does not modify the limit spectra, except for the outliers. 

\begin{theorem} 
\label{theo-boundary}
Introduce
\begin{align}
\Lambda
\;=\;
\big\{E\in\CM\,:\,|z_L(E)|=|z_{L+1}(E)|\big\} .
\label{eq-LambdaDef}
\end{align}
Suppose that $|z_{L-1}(E)|<|z_L(E)|=|z_{L+1}(E)|<|z_{L+2}(E)|$ for all but finitely many $E \in \Lambda$. Then $\Lambda$ consists of a finite number of analytic arcs. Consider the two sets $I_0=\{L+1,\ldots,2L\}$ and $I_1=\{L,L+2,\ldots, 2L\}$ and 
suppose that the functions $\hat{q}_{\IndSet_0}$ and $\hat{q}_{\IndSet_1}$ have discrete zero sets in a neighbourhood of  $\LimitSet \setminus \mathcal{F}$. Finally introduce the set of outliers depending on the perturbation $C$ by
$$
\Gamma(C)
\;=\;
\big\{E\in\CM \setminus \LimitSet  \,:\,\hat{q}_{\IndSet_0}(E)=0  \big\}
\;.
$$
Then
\begin{equation}
\label{eq-Widom}
\lim_{N\to\infty} \spec(H_N(0,0,C))
\;=\;
\Lambda\cup\Gamma(C)
\;.
\end{equation}
\end{theorem}

Now let us come to the first main new result of the paper,  corresponding to case (d) in the list above. It will be assumed throughout that $B$ is invertible. It is again necessary to start out with some notations. For $I\subset \{1,\ldots,2L\}$,  let $I^c=\{1,\ldots,2L\}\setminus I$ be the complementary set. Then the Riesz projections satisfy $\Rr_{I^c}^E= \one - {\Rr}_{I}^E$. Further let us introduce  
\begin{equation}
\label{eq-qDef}
q_{I}(E)
\;=\;
{\det}_{2L}
\big(
{\Rr}_{I}^E
\Tt^E_{\mbox{\rm\tiny bd}}
\,-\,
\Rr_{I^c}^E\big)
\;,
\qquad
E\in\CM\setminus\Ff
\;,
\end{equation}
where the transfer matrix $\Tt^E_{\mbox{\rm\tiny bd}}$ of the boundary site is defined as in \eqref{eq-TraDef}, now for $A,B,C$: 
\begin{equation}
\label{eq-TraBDDef}
\Tt^E_{\mbox{\rm\tiny bd}}
\;=\;
\begin{pmatrix}
(E\,\one-C)B^{-1} & - A \\ B^{-1} & 0 
\end{pmatrix}
\;.
\end{equation}

\begin{theorem}
\label{thm-LimitSpecIntro}
Suppose that $B$ is invertible and $A$ has rank $r=\rk(A)$. Introduce the sets
\begin{align}
\Sigma_r
&
\;=\;
\big\{E\in\CM\,:\,
\exists \;j\geq L-r+1 \;\mbox{ such that } |z_{j}(E)|=1
\big\} 
\;, 
\label{eq-SigmaRDef}
\\
\Lambda_r
&
\;=\;
{\big\{E\in\CM\,:\,|z_{L-r}(E)|=|z_{L-r+1}(E)|\geq 1 
\big\}} ,
\label{eq-LambdaRDef}
\\
\Gamma_r
& 
\;=\;
\big\{E\in\CM \setminus (\Lambda_r\cup\Sigma_r)  \,:\,
q_{I^E_{>,r}}(E)=0 
\big\}
\;,
\label{eq-SigmaDef}
\end{align}
where
\begin{equation}
\label{eq-IE>Def}
I^E_{>,r}
\;=\;\big\{j=L-r+1,\ldots,2L\,:\,|z_j(E)|>1\big\}
\;.
\end{equation}
On each of these sets one makes the further assumptions:

\begin{itemize}

\item[{\rm (i)}] For all but finitely many $E\in\Sigma_r$ there is merely one $j$ such that $|z_j(E)|=1$, namely $|z_{j-1}(E)|<1<|z_{j+1}(E)|$.  Furthermore, the functions $q_{I'_j}$ and $q_{I'_{j+1}}$ for the index sets $I'_j=\{j,\ldots,2L\}$ and $I'_{j+1}=\{j+1,\ldots,2L\}$ have discrete zero sets in a neighbourhood of $\Sigma_r \setminus \mathcal{F}$.

\item[{\rm (ii)}]  One has $|z_{L-r-1}(E)|<|z_{L-r}(E)|=|z_{L-r+1}(E)|<|z_{L-r+2}(E)|$ for all but finitely many $E \in \Lambda_r$. For $I_0=\{L-r+1,\ldots,2L\}$ and $I_1=\{L-r,L-r+2,\ldots,2L\}$ the functions $q_{I_0}$ and $q_{I_1}$ have discrete zero sets in a neighbourhood of $\LimitSet_r \setminus \mathcal{F}$.

\item[{\rm (iii)}] The function $q_{I^E_{>,r}}$ has a discrete zero set. 

\end{itemize}
Then the limit spectrum of $H_N(A,B,C)$ satisfies
\begin{equation}
\label{eq-GenResult}
\lim_{N\to\infty} \spec(H_N(A,B,C))
\;=\;
\Lambda_r\cup\Sigma_r\cup\Gamma_r
\;.
\end{equation}
\end{theorem}

Both Theorem~\ref{theo-boundary} and Theorem~\ref{thm-LimitSpecIntro} have non-triviality assumptions on the functions $\hat{q}_I$ and $q_I$ for some $I$ which already appeared in \cite{Wid1,Del,KS} and which are quite tedious to check in concrete situations. It is, moreover, possible that these non-triviality assumptions do not hold, as show examples in both \cite{Wid1} and \cite{KS}. However, these examples require some artificially produced degeneracies and generically the non-triviality assumptions are satisfied, as shown by the following second main result of this work. It is proved in Section~\ref{sec-analyticity} by computing the large $E$ asymptotics of the Riesz-projections $\Rr^E_I$. 

\begin{theorem}
\label{theo-Expansion}
For Lebesgue almost all matrix coefficients $R,T,V,A,B$ and $C$, all functions $\hat{q}_I$, with subsets $I\subset\{1,\ldots,2L\}$ of cardinality $L$, and all functions ${q}_I$, with subsets $I\subset\{1,\ldots,2L\}$, have discrete zero sets.
\end{theorem}

Next follow numerous remarks providing supplementary information on Theorems~\ref{theo-boundary} and \ref{thm-LimitSpecIntro} and outlining their proofs.

\begin{remark}
{\rm
As already stressed above, the proof of Theorem~\ref{theo-boundary} is a minor modification of the proof in \cite{KS} resulting from the perturbation $C\not =V$.  Therefore, its full proof will only be outlined in this remark. This will also allow to point out some changes of notation compared to \cite{KS}. The main tool of the proof of Theorem~\ref{theo-boundary} is the Widom-type formula for the characteristic polynomial of {$H_N(0,0,C)$}: 
\begin{equation}
{\det}_{NL}(H_N{(0,0,C)} - E\one_{NL})
\;=\; 
\sum_{|I|= L} Z_I(E)^{N} 
\,\hat{q}_I(E) 
\;,
\label{eq-OpenIdentityIntro}
\end{equation}
where the sum runs over subsets $I\subset\{1,\ldots,2L\}$ with cardinality $|I|=L$ and 
\begin{equation}
\label{eq-ZDef}
Z_I(E)
\;=\;
(-1)^{L}  {\det}_{L}(T)
\prod_{j\in I}z_j(E)
\;.
\end{equation}
The proof of \eqref{eq-OpenIdentityIntro} results from the first half of the proof of Theorem~31 in \cite{KS}. In the latter result, the index set was replaced by its complement, basically to follow Widom's notations from \cite{Wid1}, but it turns out to be more convenient (and actually more simple) to not do this here and in the generalization below. As a result, the factor $Z_I(E)$ defined in \eqref{eq-ZDef} uses the eigenvalues instead of their inverses. Based on \eqref{eq-OpenIdentityIntro}, the proof of Theorem~\ref{theo-boundary} now roughly proceeds as follows. For large $N$, the sum on the r.h.s. of \eqref{eq-OpenIdentityIntro} is dominated by the summand with the largest factor $Z_I(E)$ which due to ordering of the eigenvalues is simply given by the index set $I_0$. If its factor $\hat{q}_{I_0}(E)$ does not vanish, one concludes that in a sufficiently small  neighborhood of such $E$, $H_N{(0,0,C)}$ has no eigenvalue for $N$ sufficiently large. On the other hand, if the factor $\hat{q}_{I_0}(E)$ vanishes, Hurwitz's theorem allows to show that there must be such an eigenvalue. This precisely defines the set $\Gamma(C)$ of outliers. Further zeros of the characteristic polynomial can appear if there are two factors $Z_I(E)$ which both have the largest (same) modulus. This is possible whenever there the middle two  eigenvalues are of the same modulus, hence leading to the definition of the set $\Lambda$. The hypothesis in the theorem assures that, away from a finite number of points, there is no other factor $Z_I(E)$ of same modulus, and then in Section~4.8 of \cite{KS} it is shown that one can indeed construct an eigenfunction for a nearby complex energy, leading to a zero of the characteristic polynomial. {The finite number of points, where these arguments do not hold, can be dealt with by Cauchy's formula or by the fact that both $\Lambda$ and the limit spectra are closed (for the latter, see \cite{Bee}).} The proof of Theorem~\ref{thm-LimitSpecIntro} will   follow a similar reasoning, albeit based on a new Widom-type formula presented in Remark~\ref{rem-ProofOuline} and proved in Section~\ref{sec-LocalPerturbation}.
}
\hfill $\diamond$
\end{remark}

\begin{remark}
\label{rem-MoreSites}
{\rm 
Theorem~\ref{theo-boundary} states that the arcs $\Lambda$ constituting the continuous part of the limit spectrum are independent of the local perturbation $C$. This holds true for more general  perturbations  supported within a fixed finite distance $K$ from the two boundaries. Supposing that all off-diagonal entries are invertible, one can then define transfer matrices over the sites with a perturbation by formula \eqref{eq-TraDef}. Let $\Tt^{E}_{\mbox{\rm\tiny le},1},\ldots,\Tt^{E}_{\mbox{\rm\tiny le},K}$ and $\Tt^{E}_{\mbox{\rm\tiny ri},1},\ldots,\Tt^{E}_{\mbox{\rm\tiny ri},K}$ denote the corresponding transfer matrices. Then replacing \eqref{eq-qDef} by
$$
\hat{q}_{I}(E)
\;=\;
{\det}_L\Big(
\binom{0}{\one}^*
\Tt^E_{\mbox{\rm\tiny ri},K}\cdots \Tt^E_{\mbox{\rm\tiny ri},1}
\,{\Rr}_{I}^E
\,\Tt^E_{\mbox{\rm\tiny le},K}\cdots \Tt^E_{\mbox{\rm\tiny le},1}
\binom{\one}{0}\Big)
\;,
$$
all statements of Theorem~\ref{theo-boundary} remain valid.
\hfill $\diamond$
}
\end{remark}

The remaining remarks mainly concern Theorem~\ref{thm-LimitSpecIntro}.

\begin{remark}
{\rm
For rank $r=L$, namely both $A$ and $B$ are invertible, one has $\Sigma_L=\Sigma$ and $\Lambda_L=\emptyset$. Hence the limit spectrum consist of the periodic spectrum $\Sigma$ together with some outliers $\Gamma_L$ which depend continuously on $(A,B,C)$ because $q_{I^E_{>,r}}$ does so. More generally, note that $\Sigma_r$ and $\Lambda_r$ do not depend on the particular choice of the matrices $(A,B,C)$, but merely on the rank $r$ of $A$. With $r$ fixed, again only $\Gamma_r$ will depend on these matrices $(A,B,C)$. Similar as in Remark~\ref{rem-MoreSites}, a modification of the definition of $q_I$ allows to deal with any compactly supported perturbation. Again this does not modify $\Sigma_r$ and $\Lambda_r$. Let us finally note that in the semipermeable case with rank $r=0$ (namely $A$ vanishing and $B$ invertible), the two sets $\Sigma_0$ and $\Lambda_0$ are generically both non-empty, even though they take a somewhat more simple form. 
}
\hfill $\diamond$
\end{remark}

\begin{remark}
\label{rem-ArcsIsolatedPoints}
{\rm
The identities $|z_j(E)|=1$ and $|z_{L-r}(E)|=|z_{L-r+1}(E)|$ in \eqref{eq-SigmaRDef} and  \eqref{eq-LambdaRDef} respectively provide one real equation for the complex energy, while the two inequalities in \eqref{eq-LambdaRDef} hold on open sets. Generically, one hence expects  $\Sigma_r$ and $\Lambda_r$ to be locally one-dimensional subsets of the complex energy plane. As the functions $E\in\CM\to z_j(E)$ are locally analytic, one can hence show by the techniques of Lemmata~20 and 21 in \cite{KS} that $\Sigma_r$ and $\Lambda_r$ consist of a finite number of analytic arcs which do not have isolated points. {Let us note that the intersection $\Sigma_r\cap\Lambda_r$ contains only a finite number of points which are precisely the junctions between $\Sigma_r$ and $\Lambda_r$. In the middle plot of the numerical example in Fig.~\ref{fig-2}, these are the points in which the blue and black analytic arcs touch. In this plot there is a further $Y$-junction within $\Lambda_r$ corresponding to a point $E$ with a triple degeneracy $|z_{L-r-1}(E)|=|z_{L-r}(E)|=|z_{L-r+1}(E)|$.} Furthermore, $\Gamma_r$ is a discrete set, called the set of outliers. Let us briefly comment on the further assumptions (i), (ii) and (iii). Having respectively further unit eigenvalues or further eigenvalues of the same modulus are generically events of codimension $2$. Together with Theorem~\ref{theo-Expansion} this shows that the three conditions (i), (ii) and (iii) hold generically. 
}
\hfill $\diamond$
\end{remark}

\begin{remark}
\label{rem-ProofOuline}
{\rm
The proof of Theorem~\ref{thm-LimitSpecIntro} is based on a new Widom-type formula for the characteristic polynomial of {$H_N(A,B,C)$}: 
\begin{equation}
{\det}_{NL}(H_N{(A,B,C)} - E\one_{NL})
\;=\; 
 {\det}_{L}(B) \, 
\sum_{|I|\leq L+r} Z_I(E)^{N-1} 
\,q_I(E) 
\;,
\label{eq-IdentityIntro}
\end{equation}
where the sum runs over subsets $I\subset\{1,\ldots,2L\}$ with cardinality less than or equal to $L+r$. The restriction onto the sum results from the fact that for larger cardinality the corresponding factors $q_I(E)$ vanishes, as can readily be understood by inspecting the definition~\eqref{eq-qDef} ({\it e.g.}, for $I=\{1,\ldots,2L\}$ one has $\Rr^E=\one$ so that clearly $q_I(E)=0$ for $r<L$) . For most $E\in\CM$, the sum on the r.h.s. of \eqref{eq-IdentityIntro} is dominated by a single term. In particular, the trivial  bound $|I^E_{>,r}|\leq L+r$ implies that the set $I^E_{>,r}$ is admissible in the identity \eqref{eq-IdentityIntro}, and indeed provides the largest factor. The associated summand can nevertheless vanish because the $q$-factor vanishes, and these zeros define the set $\Gamma_r$ of outliers. The sets $\Sigma_r$ and $\Lambda_r$ result from a different situation, namely there are two summands in  \eqref{eq-IdentityIntro} for which the $Z$-factors have the same modulus. Then a suitable linear combination of these two can (and does) lead to a zero of the characteristic polynomial. For the set $\Sigma_r$ this scenario results from an eigenvalue $z_j(E)$ of unit modulus, while for $\Lambda_r$ there are two eigenvalues of same modulus (which is typically not $1$). The conditions (i) and (ii) exclude the possibility that there are three eigenvalues of same modulus {outside of a finite number of points}.
}
\hfill $\diamond$
\end{remark}

\begin{remark}
{\rm
The $j$ appearing in \eqref{eq-SigmaRDef} can be strictly larger than $L-r+1$, corresponding to situations where there are fewer eigenvalues of modulus strictly larger than $1$.  Comparing \eqref{eq-SigmaRDef} with \eqref{eq-PeriodicSpec2}, one concludes 
$$
\Sigma_r\,\subset\,\Sigma
\;,
\qquad
\Sigma_L\;=\;\Sigma
\;,
$$
justifying the choice of notation. It is also possible to write out an expression for $\Sigma_r$ that exhibits it as a topological boundary of an open set $\Omega$. The definition of this set uses the winding number $\Wind^E(H)$ of $z\in\SM^1\mapsto{\det}_L(H(z)-E\,\one)$:
\begin{align*}
\Omega_r
&
\;=\;
\{E\in\CM\,:\,\Wind^E(H)>-r\}
\\
&
\;=\;
\big\{E\in\CM\,:\,|I^E_{>,r}|<L+r\;\,\mbox{\rm and }\spec(\Tt^E)\cap\SM^1=\emptyset\big\} 
\;.
\end{align*}
In the first set the inequality $\Wind^E(H)>-r$ includes the fact that $\Wind^E(H)$ exists, namely $\spec(\Tt^E)\cap\SM^1=\emptyset$, so that the second equality then follows from the well-known identity ({\it e.g.} \cite{BS}) 
\begin{equation}
\label{eq-WindStates}
\big|\big\{j=1,\ldots,2L\,:\,|z_j(E)|>1\big\}\big|
\;=\;
L-\Wind^E
\;.
\end{equation}
Now for $E\in \Omega_r$, there are $m_E < L + r$ eigenvalues of modulus larger than $1$. Moving inside $\Omega_r$, one may add eigenvalues of modulus larger than $1$, so go to $m_E+1$ of them, unless one passes to $m_E+1=L-r$. This passage is though an eigenvalue on the unit circle $\SM^1$ and this is then precisely a point on the boundary $\partial\Omega_r$. Hence one deduces $\Sigma_r=\partial \Omega_r$.
}
\hfill $\diamond$
\end{remark}

\begin{remark}
{\rm
In the definition \eqref{eq-LambdaRDef} of $\Lambda_r$, if $|z_{L-r}(E)|=|z_{L-r+1}(E)|>1$  and $|z_{L-r-1}(E)|<1$, then there are $L+r+1$ eigenvalues of modulus larger than $1$. Then there are two possible choices for the largest allowed $Z$-factors in \eqref{eq-IdentityIntro}, namely those corresponding to the sets $I_0$ and $I_1$. On the other hand, if $|z_{L-r}(E)|=|z_{L-r+1}(E)|=1$ in  \eqref{eq-LambdaRDef}, then there are merely $L+r-1$ eigenvalues of modulus larger than $1$ (if $|z_{L-r+2}(E)|>1$). This case hence also corresponds to a point in $\Sigma_r$, albeit only one of those finite number of points described in Hypothesis (i). Hence $\Sigma_r\cap\Lambda_r$ may be non-zero, but consist of at most a finite number of points. An alternative definition of $\Lambda_r$ would require $|z_{L-r}(E)|=|z_{L-r+1}(E)|>1$, but then $\Lambda_r$ is not closed. 
}
\hfill $\diamond$
\end{remark}


\section{Numerical illustration of the main result}
\label{sec-Numerics}

\begin{figure}
\centering
\includegraphics[width=5.5cm]{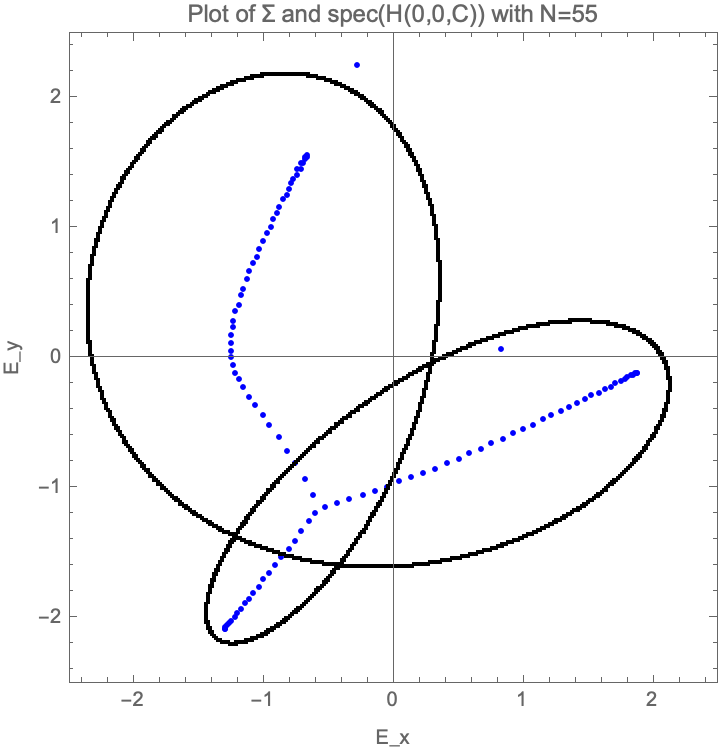}
\includegraphics[width=5.5cm]{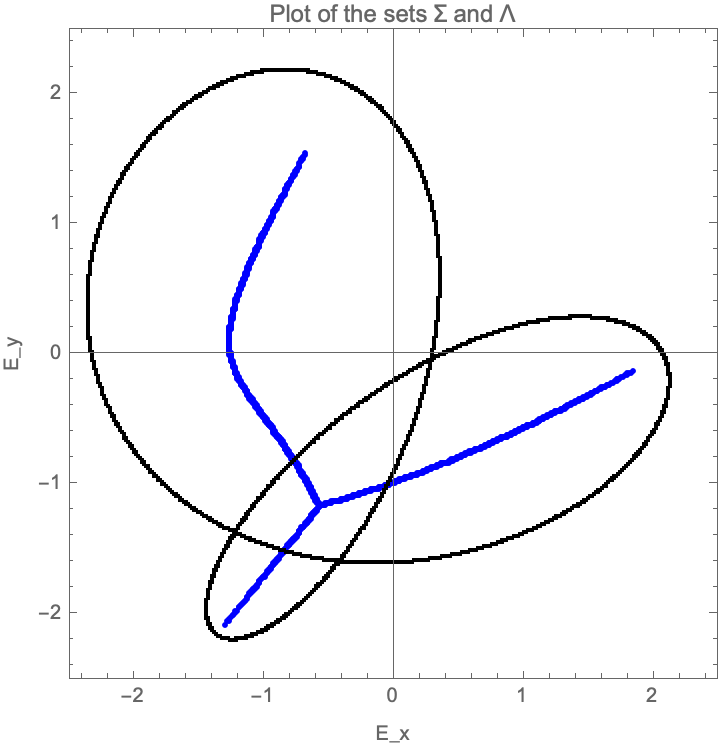}
\includegraphics[width=5.5cm]{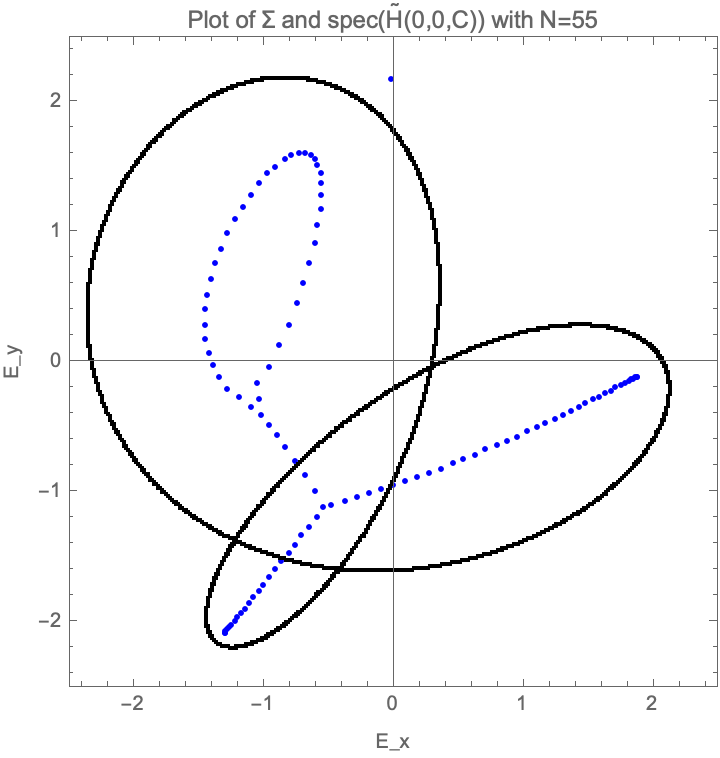}
\caption{\sl 
The first and third plots show $\spec(H_N(0,0,C))$ and $\spec(\widetilde{H}_N(0,0,C))$ as well as $\Sigma$ for $N=55$, in blue and black respectively.  The second plot shows the sets $\Lambda$ and $\Sigma$. 
}
\label{fig-1}
\end{figure}

\begin{figure}
\centering
\includegraphics[width=5.5cm]{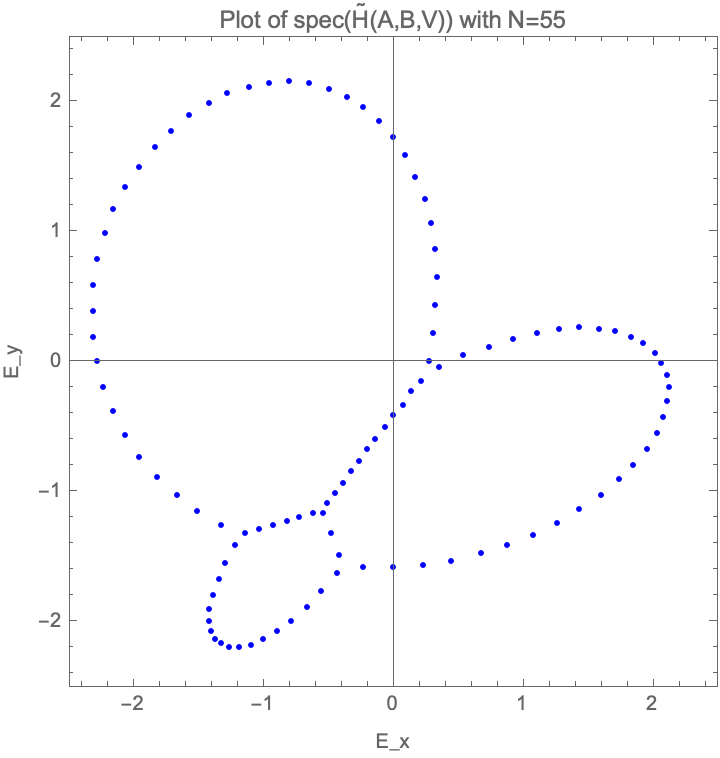}
\includegraphics[width=5.5cm]{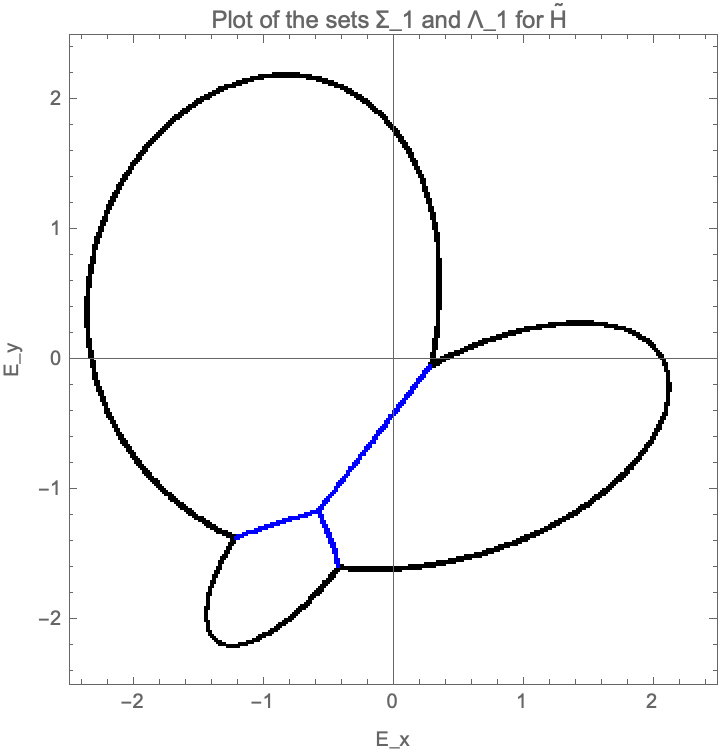}
\includegraphics[width=5.5cm]{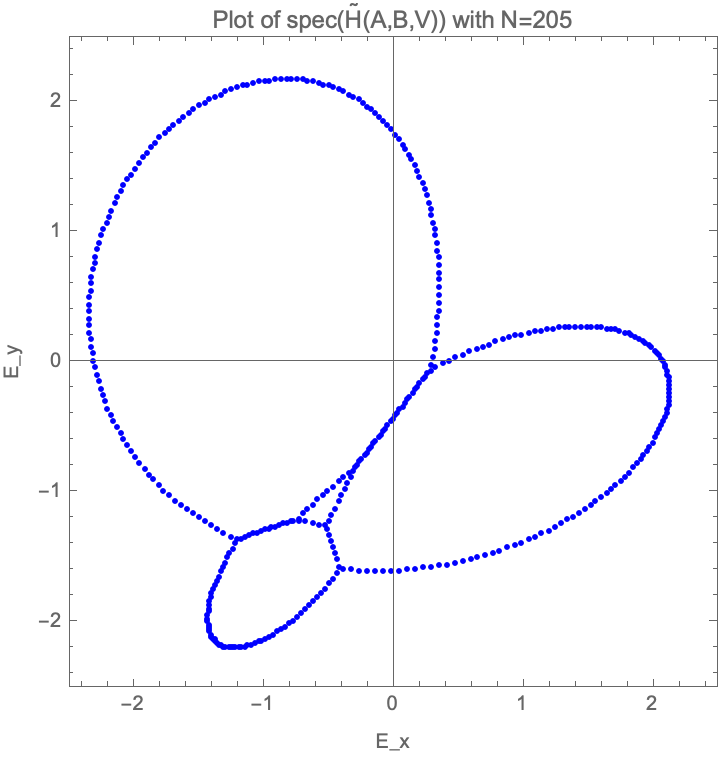}
\caption{\sl 
The first and third plots show $\spec(\widetilde{H}_N(A,B,V))$ for $N=55$ and $205$.  The second plot shows the sets $\Lambda_1$ and $\Sigma_1$, in blue and black respectively. Note that the first plot coincides with the third plot in {\rm Fig.~\ref{fig-0}} because $\spec(\widetilde{H}_N(A,B,V))=\spec(H_N(B,A,V))$.
}
\label{fig-2}
\end{figure}

\begin{figure}
\centering
\includegraphics[width=5.5cm]{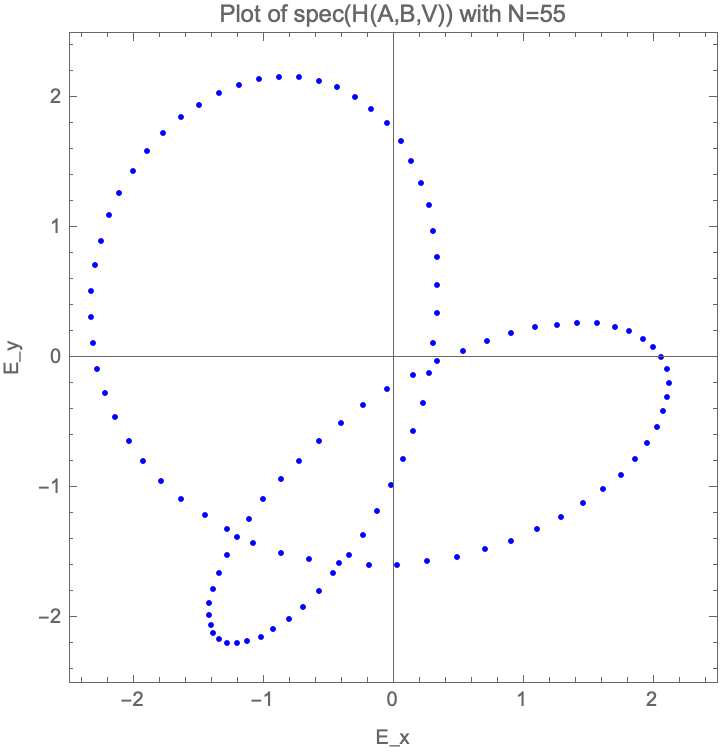}
\includegraphics[width=5.5cm]{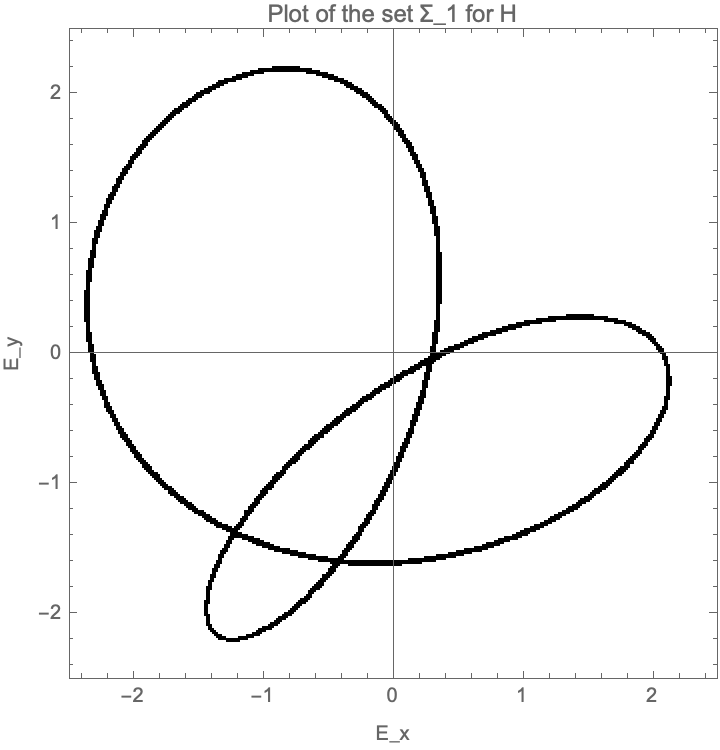}
\includegraphics[width=5.5cm]{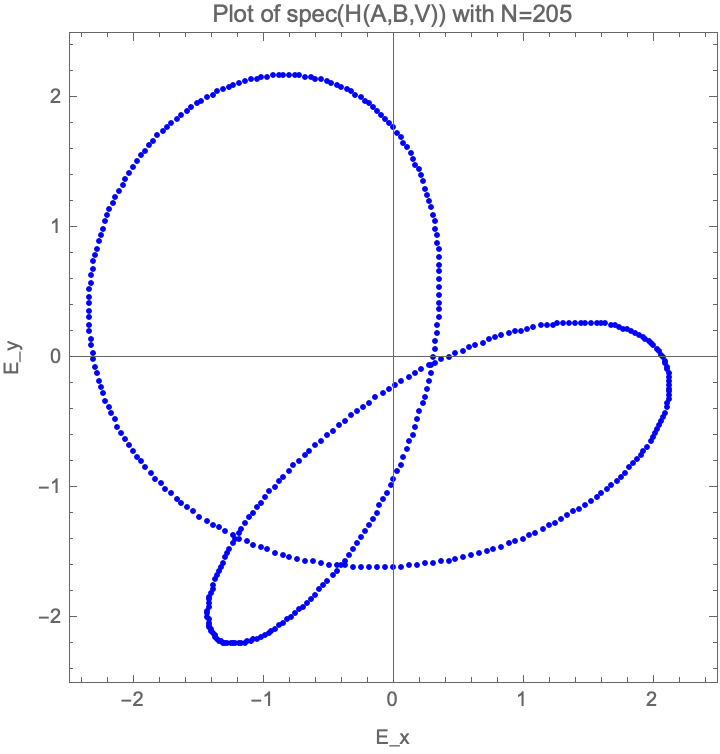}
\caption{\sl 
The first and third plots show $\spec(H_N(A,B,V))$ for $N=55$ and $N=205$. The second plot shows the set $\Sigma_1$, which coincides with $\Sigma$, while $\Lambda_1$ is empty. 
}
\label{fig-3}
\end{figure}

This section illustrates Theorems~\ref{theo-boundary} and \ref{thm-LimitSpecIntro} by some numerical results for Hamiltonians with block entries of size $L=2$. The entries of the symbol $H(z)=\frac{1}{z}R+V+zT$ are chosen to be
$$
R
\;=\;
\begin{pmatrix}
0.3\,\imath & 0.7 \\ 0 & 0.3\,\imath 
\end{pmatrix}
\;,
\qquad
T
\;=\;
\begin{pmatrix}
1.5 & 0 \\ -0.6\,\imath & 1.5 
\end{pmatrix}
\;,
\qquad
V
\;=\;
\begin{pmatrix}
0.3 -0.3\,\imath & - 0.5\,\imath \\ 1 & -0.3-0.3\,\imath 
\end{pmatrix}
\;.
$$
Apart from $H$, also the reversed symbol $\widetilde{H}(z)=\frac{1}{z}T+V+zR$ is considered. Then the associated set $\widetilde{\Sigma}$ given as in \eqref{eq-PeriodicSpec2}, clearly satisfies $\widetilde{\Sigma}=\Sigma$. However, the curve now has the opposite orientation and therefore $\Wind^E(\widetilde{H})=-\Wind^E(H)$ for all $E\in\CM\setminus\Sigma$. However, the sets $\Lambda_r$ and $\Sigma_r$ are different for $H$ and $\widetilde{H}$. For both $H$ and $\widetilde{H}$, a perturbation given by
$$
A
\;=\;
\begin{pmatrix}
0 & 1 \\ 0 & 0
\end{pmatrix}
\;,
\qquad
B
\;=\;
\begin{pmatrix}
1 & 0 \\ 0 & 0.2+\imath 
\end{pmatrix}
\;,
\qquad
C
\;=\;
\begin{pmatrix}
0.1 & -0.3 \\ 1 & 2\,\imath
\end{pmatrix}
\;.
$$
will be considered. The outcomes of numerical computations, carried out within minutes using Mathematica on a notebook, are shown in Figs.~\ref{fig-1}, \ref{fig-2} and \ref{fig-3} and will now be discussed in some detail. The blue part of the first plot in Fig.~\ref{fig-1} shows $\spec(H_N(0,0,C))$ for $N=55$. In the limit $N\to\infty$, the limit spectrum contains the set $\Lambda$, which is computed from the spectral analysis of the transfer matrix by a direct implementation of \eqref{eq-LambdaDef}, see the blue curve in the second plot of Fig.~\ref{fig-1}. The agreement is remarkably good. Also shown in both plots is the periodic spectrum $\Sigma$ which has nothing in common with $\Lambda$. Notable is also that there are two eigenvalues far off the limit curve $\Lambda$. These are precisely the zeros of {$q_{\{3,4\}}$ making up the set of outliers $\Gamma(C)$}. On the other hand, the third plot shows the spectrum of $\widetilde{H}_N(0,0,C)$ also for $N=55$. Theorem~\ref{theo-boundary} asserts that the limit spectrum should contain the same set $\Lambda$, but the plot shows a loop. This is not a finite size effect; actually the plots for larger $N$ show larger deviation from $\Lambda$. The reason for this discrepancy of theoretical results and numerical output is a numerical instability linked to the fact that typical eigenfunctions within a given region of $\CM\setminus\Sigma$ are either exponentially localized at the left or the right edge; moreover, they are very close to collinear and this leads to numerical instabilities; furthermore, all this is more pronounced for energies distant from $\Sigma$. In the physical literature on non-hermitian quantum systems \cite{KSUS,KD,OKSS} this exponential localization is called the skin effect, see the detailed discussion in \cite{KS} and references therein.  A final comment on the third plot in Fig.~\ref{fig-1} is that there is again an outlier, but at a different complex energy. This outlier is stable as the system size increases. This concludes the illustration of Theorem~\ref{theo-boundary}.

\vspace{.2cm}

Let us now come to the set-up of Theorem~\ref{thm-LimitSpecIntro}, namely consider $H_N(A,B,V)$ and $\widetilde{H}_N(A,B,V)$ with $A$ and $B$ as given above. In particular, $r=\mbox{\rm rk}(A)=1$, and $B$ is invertible. The first plot in Fig.~\ref{fig-2} shows the spectrum of $\widetilde{H}_N(A,B,V)$ with $N=55$, which is again distinct from $\Sigma$ and $\Lambda$. This fits very well with the set $\Sigma_1\cup\Lambda_1$ in the second plot, as computed by implementing \eqref{eq-SigmaRDef} and \eqref{eq-LambdaRDef}, confirming Theorem~\ref{thm-LimitSpecIntro}. However, the third plot shows that for $N=205$ agreement is again less good. This is again related to numerical instabilities linked to the skin effect. More precisely, on $\Sigma_1$ there is no skin effect and the eigenfunctions are plane waves, but on $\Lambda_1$ there is a skin effect and this is responsible for the deviations leading to a new loop in the numerically computed spectrum.

\vspace{.2cm}

Finally, Fig.~\ref{fig-3} is identical to Fig.~\ref{fig-2}, but for $H$ instead of $\widetilde{H}$.   In this case, there $\Lambda_1=\emptyset$ (except for boundary points) and $\Sigma_1=\Sigma$. As a consequence, there is no skin effect and indeed also for the large value $N=205$ there is no numerical deviation from the theoretical result. To conclude the discussion, let us explain how the opposite winding numbers lead to distinct spectral properties in Fig.~\ref{fig-2} and \ref{fig-3}. By \eqref{eq-WindStates}, the  number of eigenvalues of $\Tt^E$ of modulus less than $1$ is equal to $\Wind^E(H)-L$. For $H$ the number of eigenvalues of $\Tt^E$ inside the unit circle is equal to $3$ or $4$ in the bounded components of $\CM\setminus\Sigma$, while for $\widetilde{H}$ there are $1$ or $0$ of such eigenvalues of the associated transfer matrix $\widetilde{\Tt}^E$. In particular, in the central part where $|\Wind^E(H)|=2$, there are none and hence its boundary does not belong to $\Sigma_1$. On the other hand, in this region one may have points from $\Lambda_1$ because this involves the equal modulus of the second and third eigenvalues $z_2(E)$ and $z_3(E)$ which are indeed larger than $1$. This leads to parts of the limit spectrum shown in the second plot of Fig.~\ref{fig-2}. For $H$, there are four eigenvalues of $\Tt^E$ of modulus less $1$ in the region with $\Wind^E(H)=2$. Hence on the boundary of this region $|z_4(E)|=1$  which is hence a point in $\Sigma_1$. One concludes $\Sigma_1=\Sigma$. On the other hand, $|z_1(E)|=|z_2(E)|\geq 1$ is impossible for all $E$ in this case, so that indeed $\Lambda_1=\emptyset$.

\section{Case of periodic boundary conditions}
\label{sec-PeriodicCase}

In this section it will be shown how the framework from \cite{KS} can be applied to compute the limit of spectra of circulant (or periodic) block Toeplitz matrices $H_N(R,T,V)$, {namely $A=R$, $B=T$ and $C=V$.} As stated above, it is assumed that  $T$ and $R$ are invertible. Then $H_N{(R,T,V)}$ is block diagonalised by the unitary discrete Fourier transform $\mathcal{F}_N$ given by $(\mathcal{F}_N)_{j,k} = (2\pi)^{-\frac{1}{2}}e^{-2 \pi \imath (j-1) (k-1) N^{-1}} \one_L$ where $j,k=1,\ldots,N$. In particular, 
\begin{align*}
\mathcal{F}_N H_N{(R,T,V)} \mathcal{F}_{N}^* 
\;=\; 
\oplus_{n=1}^N\,H\big(e^{\tfrac{2\pi\imath n}{N}}\big)
\;,
\end{align*}
where as above $H(z) = z T + V + z^{-1} R$ is the symbol of $H_N$. Therefore
\begin{align*}
\spec(H_N{(R,T,V)}) 
= & \; \bigcup_{n=1}^N \spec( H(e^{\frac{2 \pi \imath n}{N}}) ) \;.
\end{align*} 
In particular, the limit spectrum is given by
\begin{align}
\label{eq-PeriodicLimitSpectrum}
\lim_{N\to\infty} \spec(H_N{(R,T,V)}) 
\;=\; 
\bigcup_{z \in \mathbb{S}^1} \spec(H(z)) \;.
\end{align}
As a preparation for the next section, we will now rederive this result using the transfer matrix approach. Let us start out with a Widom-type formula for this case, which is hence a special case of the formula in Remark~\ref{rem-ProofOuline}.

\begin{proposition}
\label{prop-CharPol}
The characteristic polynomial of $H_N{(R,T,V)}$ is given by
\begin{align}
\label{eqprop-CharPol}
{\det}_{NL}(H_N{(R,T,V)} - E\,\one_{NL}) 
\;=\; 
(-1)^{L(N-1)} {\det}_{L}(T)^N {\det}_{2L}((\mathcal{T}^E)^N - \one_{2L}) \;.
\end{align}
\end{proposition}

\noindent {\bf Proof.} For later use, let us write {out the} eigenvalue equation $H_N{(A,B,C)} \phi=E\phi$ for an eigenstate $\phi=(\phi_n)_{n=1,\ldots,N}$, $\phi_n\in\CM^L$, {in the case of} general $A$, $B$ and $C$. Entrywise it reads
\begin{align}
E\,\phi_1
& \;=\;
A\,\phi_N\,+\,C\,\phi_1\,+\,T\,\phi_2\;,
\nonumber
\\
E\,\phi_n
& \;=\;
R\,\phi_{n-1}\,+\,V\,\phi_n\,+\,T\,\phi_{n+1}\;,
\qquad n=2,\ldots,N-1\;,
\label{eq-Recurrence}
\\
E\,\phi_N
& \;=\;
R\,\phi_{N-1}\,+\,V\,\phi_N\,+\,B\,\phi_1\;.
\nonumber
\end{align} 
This can be rewritten using transfer matrices as
\begin{align}
\begin{pmatrix}T\,\phi_2 \\ \phi_1 \end{pmatrix}
& 
\;=\;
\begin{pmatrix} E\,\one - C & -A \\ \one & 0 \end{pmatrix}
\begin{pmatrix} \phi_1 \\ \phi_N \end{pmatrix}
\;,
\nonumber
\\
\begin{pmatrix}T\,\phi_{n+1} \\ \phi_n \end{pmatrix}
& 
\;=\;
\begin{pmatrix}(E\,\one-V)T^{-1} & -R \\ T^{-1} & 0 \end{pmatrix}
\begin{pmatrix}T\,\phi_n \\ \phi_{n-1} \end{pmatrix}
\;,
\qquad n=2,\ldots,N-1\;,
\label{eq-TransferForm}
\\
\begin{pmatrix}B\,\phi_{1} \\ \phi_N \end{pmatrix}
& 
\;=\;
\begin{pmatrix}(E\,\one-V)T^{-1} & -R \\ T^{-1} & 0 \end{pmatrix}
\begin{pmatrix}T\,\phi_N \\ \phi_{N-1} \end{pmatrix}
\;.
\nonumber
\end{align} 
%
%
The three equations of \eqref{eq-TransferForm}, applied in the opposite order, lead to
\begin{align}
\label{eq-TransferLeftRight}
\begin{pmatrix} B\phi_1
 \\ \phi_N \end{pmatrix}
& 
\;=\;
\Tt^E
\begin{pmatrix}T\,\phi_N \\ \phi_{N-1} \end{pmatrix}
\;=\;
(\Tt^E)^{N-1}
\begin{pmatrix}T\,\phi_2 \\ \phi_{1} \end{pmatrix}
\;=\;
(\Tt^E)^{N-1}
\begin{pmatrix}
E\,\one-C & -A \\ \one & 0 
\end{pmatrix}
\begin{pmatrix}  \phi_1 \\ \phi_N \end{pmatrix}
\;.
\end{align}
For the special case $A=R$, $B=T$ and $C=V$ considered here (corresponding to the {circulant} matrices) this becomes
$$
\begin{pmatrix} T\phi_1
 \\ \phi_N \end{pmatrix}
\;=\;
(\Tt^E)^{N}
\begin{pmatrix}  T\phi_1 \\ \phi_N \end{pmatrix}
\;.
$$
Hence a solution $\phi$ of $H_N{(R,T,V)}\phi=E\phi$ provides a vector in the kernel of $(\Tt^E)^{N}-\one_{2L}$, and inversely having such a vector in the kernel, can produce an eigenvector for $H_N{(R,T,V)}$ using the transfer matrix equations \eqref{eq-TransferForm}. Hence  the polynomials (in $E$) on the l.h.s. and r.h.s. of \eqref{eqprop-CharPol} have the same zeros. To verify the identity \eqref{eqprop-CharPol}, it will be shown that also the multiplicities of the zeros are the same. By the above argument this is actually obvious if $H_N{(R,T,V)}$ is diagonalizable, so let us first focus on this case. The characteristic polynomial of $H_N{(R,T,V)}$ is clearly of degree $NL$ and at first glance the polynomial ${\det}_{2L}((\mathcal{T}^E)^N - \one_{2L})$ seems to be of degree $2NL$ in $E$. However, it is actually also a polynomial of degree $NL$ in $E$ as will now be explained. Let $p$ be the characteristic polynomial of $(\mathcal{T}^E)^N$ and write $p(z) = \sum_{n=0}^{2L} p_n(E) z^n$. Since the entries of $\mathcal{T}^E$ are constant or linear in $E$, it follows that $\mathrm{deg}(p_n) \leq N(2L-n)$. Note that the entries of the inverse $(\mathcal{T}^E)^{-1}$ are also constant or linear in $E$. Hence, if $q(z) = \sum_{n=0}^{2L} q_n(E) z^n$ is the characteristic polynomial of $(\mathcal{T}^E)^{-N}$, one similarly has $\mathrm{deg}(q_n) \leq N(2L-n)$. Now note that $p(z) = {\det}_{2L}((\mathcal{T}^E)^N) z^{2L} q(\frac{1}{z})$, so that $p_n = {\det}_{2L}((\mathcal{T}^E)^N) q_{2L-n}$. It follows that $\mathrm{deg}(p_n) \leq \min \{N(2L-n), Nn\}$. Moreover, $p_L(E) = (-1)^L {\det}_L (T^{-N}) E^{NL} (1 + \Oo(E^{-1}))$. Hence ${\det}_{2L}((\mathcal{T}^E)^N - \one_{2L}) = p(1) = \sum_{n=0}^{2L} p_n(E)$ is indeed a polynomial of degree $NL$. Since the characteristic polynomial of $H_N{(R,T,V)}$ has the same zeros, counted with multiplicities, and the same degree as ${\det}_{2L}((\mathcal{T}^E)^N - \one_{2L})$ the identity \eqref{eqprop-CharPol} follows by comparing the leading coefficients of both polynomials, albeit only in the situation where $H_N{(R,T,V)}$ diagonalizable. However, by Proposition~\ref{prop-density} the set of diagonalizable circulant Toeplitz matrices is dense in the set of all circulant Toeplitz matrices. By the continuity of both sides of \eqref{eqprop-CharPol} in $R$, $T$ and $V$ one concludes that the identity also holds for non-diagonalizable $H_N{(R,T,V)}$.
\hfill $\Box$

\begin{remark}
{\rm
The proof of Proposition~\ref{prop-CharPol} fixes an oversight in the proof of Theorem 31, more precisely the equality (70), in \cite{KS} which only works for diagonalizable $H_N{(0,0,V)}$. However, arguing again with a suitable analogue of Proposition~\ref{prop-density} this is actually sufficient.
}
\hfill $\diamond$
\end{remark}

Let us rewrite \eqref{eqprop-CharPol} as follows: 
\begin{align}
\frac{{\det}_{NL}(H_N{(R,T,V)} - E\,\one_{NL})}{(-1)^{L(N-1)} {\det}_{L}(T)^N} 
\;=\; 
 {\det}_{2L}\left( \begin{pmatrix} \one_{2L} \\ \one_{2L} \end{pmatrix}^* 
\begin{pmatrix} \mathcal{T}^E & 0 \\ 0 & \one_{2L} \end{pmatrix}^N 
\begin{pmatrix} \one_{2L} \\ -\one_{2L} \end{pmatrix}
\right) \;.
\label{eq-CharPolynomial}
\end{align}
Now we would like to apply Theorem 31 and Propositions 39 and 44 in \cite{KS} using the transfer matrix $\mathrm{diag}(\mathcal{T}^E , \one_{2L})$ to obtain the limit spectrum of $H_N{(R,T,V)}$. Note, however, that there are two things that are different in the current setting compared with the setting of \cite{KS}. Namely, here the matrices, with which the transfer matrix is multiplied and encode the boundary, are different and the transfer matrix now has eigenvalues with algebraic multiplicity higher than one for all $E$, ({\it i.e.} one is in the case $\mathcal{F} = \mathbb{C}$ in \cite{KS}.)

\vspace{.2cm}

Let the eigenvalues of $\Tt^E$ be denoted by $z_j(E)$, $j=1,\ldots,2L$ ordered according to their modulus, namely $|z_j(E)|\leq|z_{j+1}(E)|$, and the associated Riesz projection of $\Tt^E$ by $\Rr^E_j$. Suppose the $E\not\in\Ff$, namely all eigenvalues are simple. Then
$$
\Tt^E
\;=\;
\sum_{j=1}^{2L} z_j(E)\,\Rr^E_j
\;,
$$
is a decomposition of $\Tt^E$ into a sum of rank-one operators. Using the same Riesz projections also for a decomposition of the identity $\one_{2L}$, one also gets
$$
\begin{pmatrix}
\Tt^E & 0 \\ 0 & \one_{2L}
\end{pmatrix}
\;=\;
\sum_{j=1}^{2L} z_j(E)\,
\begin{pmatrix}
\Rr^E_j & 0 \\ 0 & 0
\end{pmatrix}
\;+\;
\sum_{j=1}^{2L} 
\begin{pmatrix}
0 & 0 \\ 0 & \Rr^E_j
\end{pmatrix}
\;,
$$
which is again a decomposition of the matrix on the l.h.s. into rank-one operators. This decomposition is used to compute the {product} on the r.h.s. of \eqref{eq-CharPolynomial}. Indeed, {similar as in} \cite[Lemma~6.3]{Wid1} (see also \cite[Lemma 32]{KS}) one gets
\begin{align*}
{\det}_{NL}(H_N{(R,T,V)} - E\,\one_{NL})
\;=\; 
(-1)^L
\sum_{I,J} Z_I(E)^N 
{\det}_{2L}
\left( \begin{pmatrix} \one_{2L} \\ \one_{2L} \end{pmatrix}^* 
\begin{pmatrix} \Rr^E_I & 0 \\ 0 & \Rr^E_J \end{pmatrix}
\begin{pmatrix} \one_{2L} \\ -\one_{2L} \end{pmatrix}
\right) \;,
\end{align*}
where the sum runs over subsets $I, J \subset \{1,\ldots,2L\}$ satisfying $|I|+|J|=2L$, and $Z_I(E)$ is defined as in \eqref{eq-ZDef} together with $Z_\emptyset(E)=1$. Multiplying out the factors in the determinant gives
$$
{\det}_{2L}
\left( \begin{pmatrix} \one_{2L} \\ \one_{2L} \end{pmatrix}^* 
\begin{pmatrix} \Rr^E_I & 0 \\ 0 & \Rr^E_J \end{pmatrix}
\begin{pmatrix} \one_{2L} \\ -\one_{2L} \end{pmatrix}
\right) 
\;=\;
{\det}_{2L}(\Rr^E_I - \Rr^E_J)
\;.
$$
This determinant vanishes for all choices of $I$ and $J$ except if $I\cup J=\{1,\ldots,2L\}$, namely $J=I^c=\{1,\ldots,2L\}\setminus I$. In this latter case, ${\det}_{2L}(\Rr^E_I - \Rr^E_{I^c})=(-1)^{|I|}$. Hence
\begin{align*}
{\det}_{NL}(H_N{(R,T,V)} - E\,\one_{NL})
\;=\; 
\sum_{I} (-1)^{L+|I|}\,Z_I(E)^N 
\;,
\end{align*}
where now the sum runs over all subsets $I\subset\{1,\ldots,2L\}$, including $I=\emptyset$. A largest summand is given by $Z_{I^E_{>}}(E)$ where 
\begin{equation}
\label{eq-IEDef}
I^E_{>}
\,=\,
\{j\;:\; |z_j(E)|>1\}
\;,
\end{equation}
which includes the case $I^E_{>}=\emptyset$ with a largest summand equal to $1$. Now, if there is no eigenvalue of modulus $1$, then $|Z_{I}(E)|<|Z_{I^E_{>}}(E)|$ for all other subsets $I$. In the limit $N\to\infty$ there can then not exist a zero of the r.h.s. On the other hand, if $\Tt^E$ has at least one eigenvalue $|z_j(E)|=1$, then for $I = {I^E_{>}} \cup \{j\}$ one has $|Z_{I}(E)|=|Z_{I^E_{>}}(E)|$ and using these two summands, one can construct eigenfunctions of ${H_N(R,T,V)}$ with eigenvalues in a vicinity of $E$, see Section~4.8 in \cite{KS}. This allows to provide an alternative proof of \eqref{eq-PeriodicLimitSpectrum}, by transposing the arguments in \cite{KS}, see the next section which deals with a more general case.

\section{Proof of Theorem~\ref{thm-LimitSpecIntro}}
\label{sec-LocalPerturbation}

\noindent {\bf Proof of Theorem~\ref{thm-LimitSpecIntro}.} Throughout, $B$ is invertible so that $\Tt^E_{\mbox{\rm\tiny bd}}$ is well-defined, see \eqref{eq-TraBDDef} for its definition. Then the proof of Proposition~\ref{prop-CharPol} based on \eqref{eq-TransferLeftRight} directly transposes to this more general case and leads to
\begin{align*}
{\det}_{NL}(H_N {(A,B,C)} - E\one_{NL}) 
\;=\; 
(-1)^{L(N-1)}{\det}_{L}(B) {\det}_{L}(T)^{N-1} {\det}_{2L}( (\mathcal{T}^E)^{N-1}\Tt^E_{\mbox{\rm\tiny bd}} - \one_{2L}) \;,
\end{align*}
Rewriting gives
\begin{align*}
\frac{{\det}_{NL}(H_N{(A,B,C)} - E\one_{NL}) }{(-1)^{L(N-1)} {\det}_{L}(B)  {\det}_{L}(T)^{N-1} }
\;=\; 
{\det}_{2L}\left(  
\begin{pmatrix} \one_{2L} \\ \one_{2L} \end{pmatrix}^*
\begin{pmatrix} \mathcal{T}^E & 0 \\ 0 & \one_{2L} \end{pmatrix}^{N-1} 
\begin{pmatrix}  \Tt^E_{\mbox{\rm\tiny bd}} \\ -\one_{2L} \end{pmatrix}
\right) \;.
\end{align*}
Proceeding further as in Section~\ref{sec-PeriodicCase} leads to 
\begin{align*}
{\det}_{NL}(H_N{(A,B,C)} - E\one_{NL})   
&
\;=\; 
{\det}_{L}(B) 
\sum_{I,J} Z_I(E)^{N-1} 
{\det}_{2L}\left(
\Rr^E_I \Tt^E_{\mbox{\rm\tiny bd}}-\Rr^E_J
\right) \;,
\end{align*}
where the sum runs over subsets $I\subset \{1,\ldots,2L\}$ and $J\subset \{1,\ldots,2L\}$ satisfying $|I|+|J|=2L$. Again merely the summands with $J=I^c$ can contribute because otherwise the rank of $\Rr^E_I \Tt^E_{\mbox{\rm\tiny bd}}-\Rr^E_J$ is less than $2L$. Thus
\begin{align*}
{\det}_{NL}(H_N{(A,B,C)} - E\one_{NL}) 
&
\;=\; 
{\det}_{L}(B) 
\sum_{I} Z_I(E)^{N-1} 
{\det}_{2L}\left(
\Rr^E_I \Tt^E_{\mbox{\rm\tiny bd}}-\Rr^E_{I^c}
\right) \;,
\end{align*}
with a sum running over all subsets $I\subset \{1,\ldots,2L\}$. However, now not all the determinants on the r.h.s. are necessarily non-vanishing. Indeed, the sum can be restricted to those $I$ for which $\Ran(\Rr^E_I\Tt^E_{\mbox{\rm\tiny bd}})=\Ran(\Rr^E_{I})$. Next note that
$$
\rk(\Rr^E_I\Tt^E_{\mbox{\rm\tiny bd}})
\;\leq\;
\min\big\{\rk(\Rr^E_I),\rk(\Tt^E_{\mbox{\rm\tiny bd}})\big\}
\;=\;\min\big\{|I|,L+r\big\}
\;.
$$
Thus, if $L+r<|I|$, then $\Ran(\Rr^E_I\Tt^E_{\mbox{\rm\tiny bd}})\not=\Ran(\Rr^E_{I})$ and the corresponding summand vanishes. Hence, taking into account the definition \eqref{eq-qDef}, one deduces the generalized Widom formula \eqref{eq-IdentityIntro} already discussed in the overview:
\begin{align}
{\det}_{NL}(H_N{(A,B,C)} - E\one_{NL})
&
\;=\; 
{\det}_{L}(B)
\sum_{I,\;|I|\leq L+r} Z_I(E)^{N-1} 
\,q_I(E)
\;.
\label{eq-Identity}
\end{align}
The maps $E\in\CM\setminus\Ff\mapsto q_I(E)$ are locally analytic ({\it i.e.} analytic in neighborhoods of any point in $\CM\setminus\Ff$, but this does {\it not} mean that it is well-defined on a pointed neighborhood of a point in $\Ff$). Hence each $q_I$ is either identically zero or its zeros form a discrete set. In the following, the focus will mainly be on two summands in \eqref{eq-Identity}, and for those the hypotheses of Theorem~\ref{thm-LimitSpecIntro} will assure that one is in the latter case.

\vspace{.1cm}

On the r.h.s. of \eqref{eq-Identity}, there is the summand with $I^E_{>,r}$ defined in \eqref{eq-IE>Def}, because $|I^E_{>,r}|\leq L+r$. The associated factor $|Z_{I^E_{>,r}}(E)|$ is larger than or equal to all other factors $|Z_I(E)|$. If $|Z_{I^E_{>,r}}(E)|>|Z_I(E)|$ for all $I\not=I^E_{>,r}$ and its factor in \eqref{eq-Identity}, given by $q_{I^E_{>,r}}(E)$, does not vanish, then the locally uniformly convergent sequence on the r.h.s. of \eqref{eq-Identity} does not have a zero for $N$ sufficiently large. It follows that $E$ is {\it not} in {$\limsup_{N\to\infty} \spec(H_N(A,B,C))$}. Due to the definitions of the sets $\Lambda_r$, $\Sigma_r$ and $\Gamma_r$, this implies that $E\not\in \Lambda_r\cup\Sigma_r\cup\Gamma_r$ does not lie in {$\limsup_{N\to\infty} \spec(H_N(A,B,C))$}. This shows the inclusion $\subset$ in \eqref{eq-GenResult}, {modulo the fact that superior and inferior limits coincide.}

\vspace{.1cm}

Let now $E\in\Gamma_r {\setminus\Ff}$. Then still $Z_{I^E_{>,r}}(E)$ is the largest factor in \eqref{eq-Identity} compared to the other $Z_{I}(E)$, but it is multiplied by $q_{I^E_{>,r}}(E)$ which vanishes at $E\in\Gamma_r$. However, slightly away from $E\in\Gamma_r$ it does not vanish by assumption (iii), and then again the corresponding summand dominates all others for $N$ sufficiently large. Due to Hurwitz's theorem the locally uniformly convergent series does have a zero in a neighborhood of $E\in \Gamma_r$ (which is actually exponentially small in $N$). {If $E \in \Gamma_r \cap \Ff$, then one can apply Cauchy's formula as in the proof of Proposition 39 in \cite{KS} to arrive at the same conclusion.} This zero then lies in the spectrum $\spec(H_N(A,B,C))$, and one concludes that $\Gamma_r$ is a subset of the {inferior limit spectrum $\liminf_{N\to\infty} \spec(H_N(A,B,C))$.}

\vspace{.1cm}

Next let $E\in \Sigma_r$. Recall the definition \eqref{eq-SigmaRDef} of $\Sigma_r$. By Hypothesis (i), one can, moreover, use $|z_{j-1}(E)|<1<|z_{j+1}(E)|$. Therefore there are two allowed $Z$-factors of the same and largest modulus, namely those associated with $I'_j=\{j,\ldots,2L\}$ and $I'_{j+1}=\{j+1,\ldots,2L\}$ (note that they indeed satisfy $|I'_{j+1}|<|I'_j|\leq L+r$). As $|Z_{I'_j}(E)|=|Z_{I'_{j+1}}(E)|$ and all other $Z_I(E)$ have strictly smaller modulus, it is conceivable that a linear combination of these two summands in \eqref{eq-Identity} leads to a zero of the characteristic polynomial for some $E'$ close to $E$. Using the second part of Hypothesis (i), this can indeed be achieved by constructing the corresponding eigenfunction of $H_N(A,B,C)$ explicitly by the argument in Section~4.8 in \cite{KS}. This implies that $E\in \Sigma_r$ lies in the {inferior} limit spectrum. For a discrete set of $E\in \Sigma_r$ the condition $|z_{j-1}(E)|<1$ may not hold. But as $\Sigma_r$ has no isolated points (see Remark~\ref{rem-ArcsIsolatedPoints}) and is closed on the one hand, and also the {inferior} limit spectrum is closed on the other hand {(actually both inferior and superior limit sets are closed sets \cite[Section~5.2]{Bee})}, one concludes that all $\Sigma_r$ lies in the {inferior} limit spectrum. 

\vspace{.1cm}

Finally let us consider some $E\in\Lambda_r$. Then the two sets $I_0$ and $I_1$ appearing in the statement of Theorem~\ref{thm-LimitSpecIntro} satisfy by definition $|I_0|=|I_1|=L+r$ and $|Z_{I_0}(E)|=|Z_{I_{1}}(E)|$. Hypothesis (ii) implies that all other $Z$-factors are of strictly smaller modulus, except for a discrete set of points in $\Lambda_r$. Furthermore, the second part of Hypothesis (ii) assures that again the argument of Section~4.8 in \cite{KS} can be applied to construct an eigenfunction of $H_N(A,B,C)$ for an eigenvalue in a neighborhood of $E$ (of size that is exponentially small in $N$). As $\Lambda_r$ is closed and contains no isolated points (see again Remark~\ref{rem-ArcsIsolatedPoints}), this implies as above that $\Lambda_r$ is contained in the {inferior} limit spectrum.

\vspace{.1cm}

{In conclusion, the above shows
$$
\Lambda_r\cup\Sigma_r\cup\Gamma_r
\,\subset\,
\liminf_{N\to\infty} \spec(H_N(A,B,C))
\,\subset\,
\limsup_{N\to\infty} \spec(H_N(A,B,C))
\,\subset\,
\Lambda_r\cup\Sigma_r\cup\Gamma_r
\;,
$$
which implies the claim \eqref{eq-GenResult}.}
\hfill $\Box$

\section{Asymptotics of the $\hat{q}$- and $q$-functions}
\label{sec-analyticity}

This section provides the proof of Theorem~\ref{theo-Expansion}. This requires to control the asymptotics of the functions $\hat{q}_I$ and $q_I$ for any index set $I\subset\{1,\ldots,2L\}$. As first step, the next lemma determines the asymptotic behavior of the Riesz projections for large $E$.

\begin{proposition}
\label{prop-RieszExpand1}
Assume that $T$ and $R$ have simple spectrum. Write $r_i$ and $t_i$ for the eigenvalues of $R$ and $T$ respectively, labeled such that $|r_1| < \dots < |r_L|$ and $|t_L| < \dots < |t_1|$. Furthermore let us label the associated Riesz projections as follows: for $i=1,\ldots,L$, let $P^R_{i}$ be the Riesz projection of $R$ associated to eigenvalue $r_i$ and set $P^T_{i}=0$,  while for $i=L+1,\ldots,2L$ set $P^R_{i}=0$ and let $P^T_{i}$ be the Riesz projection of $T$ associated to eigenvalue {$t_{i-L}$}. Finally set
$$
P^T_I \,=\, \sum_{i \in I} P^T_{i} 
\quad \text{ and } \quad 
P^R_I \,=\, \sum_{i \in I} P^R_{i}
\;.
$$
Then the Riesz projections $\mathcal{R}_I^E$ are analytic in $E$ at infinity with expansion given by
\begin{equation}
\label{eq-RieszExpansion}
\mathcal{R}_I^E 
\;=\; 
\begin{pmatrix} P^T_I + \frac{1}{E} Q^T_I & \frac{1}{E} T (P^R_I - P^T_I ) R \\ 
- \frac{1}{E} (P^R_I - P^T_I ) & P^R_I + \frac{1}{E} Q^R_I \end{pmatrix} 
\;+\; 
\mathcal{O}(E^{-2}) \;,
\end{equation}
where $Q^T_I$ and $Q^R_I$ are some matrices that are off-diagonal w.r.t. $P^T_I$ and $P^R_I$ respectively, i.e.
$$
P^T_I Q^T_I P^T_I \,=  \, ( \one_L - P^T_I) Q^T_I ( \one_L - P^T_I) \,= \,0 \;,
\qquad
P^R_I Q^R_I P^R_I\, =\,  ( \one_L - P^R_I) Q^R_I ( \one_L - P^R_I) \,=\, 0 \;.
$$
\end{proposition}

Let us note that  the above somewhat intricate definitions imply that
$$
\rk(P^R_I)
\,=\,
\big|I\cap\{1,\ldots,L\}\big|
\;,
\qquad
\rk(P^T_I)
\,=\,
\big|I\cap\{L+1,\ldots,2L\}\big|
\;,
$$
so that
\begin{equation}
\label{eq-RankSum}
\rk(P^R_I)\,+\,\rk(P^T_I)\;=\;|I|
\;.
\end{equation}
%

\begin{remark}
\label{rem-LabelingAmbiguity}
{\rm
The strict inequalities $|r_1| < \dots < |r_L|$ and $|t_L| < \dots < |t_1|$ required in Proposition~\ref{prop-RieszExpand1} imply that  the set 
\begin{equation}
\label{eq-AmbiguitySet}
\mathcal{A} 
\;=\;
\big\{ E \in \CM \,:\, \exists\; i \neq j \colon |z_i(E)| = |z_j(E)| \big\}
\end{equation}
is bounded. If one only supposes $|r_1| \leq \dots \leq |r_L|$ and $|t_L| \leq \dots \leq |t_1|$, then $\Aa$ may be unbounded and in this case the labeling of the $\mathcal{R}_I^E$ in Proposition~\ref{prop-RieszExpand1} is ambiguous. As a consequence, the functions $\mathcal{R}_I^E$ might not be analytic anymore due to a relabeling. One can, however, obtain well-defined analytic functions by working with analytic continuations of the $\mathcal{R}_I^E$ through the set $\mathcal{A}$, which are formed by appropriate glueing of $\mathcal{R}_I^E$ for different $I$. Then Proposition~\ref{prop-RieszExpand1} remains valid. The same holds true for the results on the functions $\tilde{q}_I$, $\hat{q}_I$ and $q_I$ in Propositions \ref{prop-qtildeAsymp}, \ref{prop-qhatAsymp} and \ref{prop-qAsymp} below. 
}
\hfill $\diamond$
\end{remark}

\noindent {\bf Proof} of Proposition~\ref{prop-RieszExpand1}.
First of all, the analyticity of the $\mathcal{R}_I^E$ will be shown. These are the Riesz projections of $\Tt^E$ and thus also of
\begin{equation}
\label{eq-TransferMatrixInverseExpansion}
\frac{1}{E} \,\Tt^E 
\;=\; 
\begin{pmatrix} T^{-1} & 0 \\ 0 & 0 \end{pmatrix}\, +\, \mathcal{O}(E^{-1}) 
\qquad \text{and} \qquad 
\frac{1}{E} \,(\Tt^E)^{-1} 
\;=\; 
\begin{pmatrix} 0 & 0 \\ 0 & R^{-1} \end{pmatrix} \,+\, \mathcal{O}(E^{-1})
\;.
\end{equation}
Since $R$ and $T$ have simple spectrum, it follows from analytic perturbation theory that the $\mathcal{R}_i^E$ are analytic at infinity for all $i$. Since the $\mathcal{R}_I^E$ are just sums of the $\mathcal{R}_i^E$, it follows that the $\mathcal{R}_I^E$ are indeed analytic in $E$ at infinity for all $I$. Second, the expansion coefficients will be computed. Let us start out with the identity
$$
\mathcal{R}_I^E 
\;=\; 
\begin{pmatrix} T & 0 \\ 0 & \one \end{pmatrix} 
\intop_{\gamma_I} \frac{dz}{2\pi \imath} \begin{pmatrix} (H(z)-E)^{-1} & -z^{-1} (H(z)-E)^{-1} \\ z^{-1 }(H(z)-E)^{-1} & z^{-1}R^{-1} - z^{-2}(H(z)-E)^{-1} \end{pmatrix} 
\begin{pmatrix} \one & 0 \\ 0 & R \end{pmatrix} \;. 
$$
This follows from \eqref{eq-RI} when the resolvent of the transfer matrix as given by the Schur complement formula is replaced, see equation (19) in \cite{KS}. Note that the paths $\gamma_I$ indeed exists for all $I$ at large $E$ since $\Ff$ is finite. Hence one needs to compute the residues of the functions $z \mapsto z^{-k}(H(z)-E)^{-1}$. Let us write $E_j(z)$ for the eigenvalues of $H(z)$ and $P_j(z)$ for the associated spectral projections. Since $H(z) = zT + V + z^{-1}R$, it follows by analytic perturbation theory that
\begin{align*}
E_j(z) \;= & \; \frac{r_j}{z} \,+\, \mathcal{O}(1) \;, &  P_j(z)\; = & \; P^R_{j} \,+\, \mathcal{O}(z )\,, & & \text{around } z = 0\;, \\
E_j(z)\; = & \; t_j z \,+\, \mathcal{O}(1) \;, & P_j(z) \;= & \; P^T_{j+L} \,+\, \mathcal{O}(z^{-1})\,, & & \text{around } z = \infty \;.
\end{align*}
Now note that  
$$
\prod_{j=1}^L (E_j(z) - E) 
\;=\; 
\det\!_{L}(H(z)-E) 
\;=\; 
\frac{\det\!_{L}(T)}{z^L}\, \det\!_{2L}(\mathcal{T}^E - z) 
\;=\; 
\frac{\det\!_{L}(T)}{z^L} \prod_{i=1}^{2L} (z- z_i(E)) 
\;, 
$$
where the $z_i(E)$ are the eigenvalues of the transfer matrix $\mathcal{T}^E$. Due to \eqref{eq-TransferMatrixInverseExpansion} one has the following expansions around $E = \infty$ for $1 \leq i \leq L$ (see also the proof of Lemma 24 in \cite{KS})
$$
z_i(E)
\;=\; 
\frac{r_i}{E} \,+\, \mathcal{O}(E^{-2}) 
\quad \text{and} \quad 
z_{L+i}(E) \;=\;\frac{E}{t_i} \,+\, \mathcal{O}(1) \;.
$$
It follows from the above and the fact that $z_i(E) \to 0$ and $z_{L+i}(E) \to \infty$ as $E \to \infty$, that for large $E$
$$
E_i(z_i(E)) \;=\; E_i(z_{L+i}(E)) \;= \;E
\;.
$$
To compute the residues let us consider the limits
$$
\lim_{z\to z_i(E)} \frac{z-z_i(E)}{E_j(z) - E} \;.
$$
First suppose that $j \neq i$. Then, for large $E$, $z_i(E)$ is small and
$$
E_j(z_i(E)) - E 
\;=\; 
\frac{r_j}{z_i(E)} \,-\, E \,+\, \mathcal{O}\big((z_i(E))^0\big) 
\;=\; 
\Big(\frac{r_j}{r_i} - 1\Big) E \,+\, \mathcal{O}(E^0) \;,
$$
so that $E_j(z_i(E)) - E$ is non-zero for large $E$. Similarly, $E_j(z_{L+i}(E)) - E$ is non-zero for large $E$, because $z_{L+i}(E)$ is big and
$$
E_j(z_{L+i}(E)) - E 
\;=\; 
\frac{z_{L+i}(E)}{t_j} \,-\, E + \mathcal{O}\big((z_{L+i}(E))^0\big) 
\;=\; 
\Big(\frac{t_i}{t_j} - 1\Big) E \,+\, \mathcal{O}(E^0) 
\;.
$$
It follows that, still for large $E$,
$$
\lim_{z\to z_i(E)} \frac{z-z_i(E)}{E_j(z) - E} 
\;=\; 
0 
\;=\; 
\lim_{z\to z_{L+i}(E)} \frac{z-z_{L+i}(E)}{E_j(z) - E} 
\;.
$$
For the case $j=i$ one argues as follows. Since $H(z)$ has simple spectrum for $z$ small or large enough (because $R$ and $T$ have simple spectrum), it follows by analytic perturbation theory that the scaled eigenvalues $z E_j(z)$ and $z^{-1} E_j(z)$ of $H(z)$ are holomorphic around $0$ and $\infty$ respectively, say for all $z$ such that $|z|<\rho_R$ and $|z|>\rho_T$ respectively. Now, if $E$ is large enough, then $|z_i(E)|<\rho_R$ and $|z_{L+i}(E)|>\rho_T$, so $E_i$ is holomorphic in the points $z_i(E)$ and $z_{L+i}(E)$. Next let us expand  $z\mapsto E_i(z)$ once around $0$ and once around $z_i(E)$:
$$
E_i(z) 
\;=\;
\frac{r_i}{z} \,+\, \sum_{n=0}^\infty a_n z^{n} 
\;=\; 
\sum_{n=0}^\infty b_n (z - z_i(E))^n 
\;,
$$
where $a_n,b_n\in\CM$ are some expansion coefficients. Recall that $E_i(z_i(E)) = E$, and thus $b_0 = E$. By differentiating both sums and evaluating in $z_{i}(E)$, one obtains
$$
b_1 
\;=\; 
- \frac{r_i}{(z_{i}(E))^2} \,+\, \sum_{n=1}^\infty n a_n (z_{i}(E))^{n-1} 
\;=\; 
-\frac{E^2}{r_i} 
\,+\, 
\mathcal{O}(E) \;. 
$$
This gives
$$
\lim_{z\to z_i(E)} \frac{z-z_i(E)}{E_i(z) - E} 
\;=\; 
\frac{1}{b_1} 
\;=\; 
-\frac{r_i}{E^2} \,+\, \mathcal{O}(E^{-3}) 
\;.
$$
Similarly, when comparing the series expansion of $E_i$ around infinity and $z_{L+i}(E)$ one obtains
$$
\lim_{z\to z_i(E)} \frac{z-z_i(E)}{E_i(z) - E} 
\;=\; 
\frac{1}{t_i} \,+\, \mathcal{O}(E^{-2}) \;.
$$
Finally one can compute the residues. Since at large $E$ the transfer matrix has simple spectrum, it follows that the residues for large $E$ are given by
\begin{align*}
\mathrm{Res}_{z_{i}(E)}(z^{-k}(H(z) - E)^{-1}) 
\; = & \; \lim_{z\to z_{i}(E)} \frac{z-z_{i}(E)}{z^k} (H(z) - E)^{-1} \\ 
= & \; \lim_{z\to z_{i}(E)} \frac{1}{z^k} \sum_{j=1}^L \frac{z-z_{i}(E)}{E_j(z) - E} P_j(z) \\ 
= & \; \frac{1}{(z_{i}(E))^k} \Big(-\frac{r_i}{E^2}\, +\, \mathcal{O}(E^{-3})\Big) P_i(z_{i}(E)) \\ 
= & \; - \frac{r_i^{1-k}}{E^{2-k}} \,P^R_{i}\, +\, \mathcal{O}(E^{k-3}) 
\;,
\end{align*}
and
\begin{align*}
\mathrm{Res}_{z_{L+i}(E)}(z^{-k}(H(z) - E)^{-1}) 
= & \; \lim_{z\to z_{L+i}(E)} \frac{z-z_{L+i}(E)}{z^k} (H(z) - E)^{-1} \\ 
= & \; \lim_{z\to z_{L+i}(E)} \frac{1}{z^k} \sum_{j=1}^L \frac{z-z_{L+i}(E)}{E_j(z) - E} P_j(z) \\ 
= & \; \frac{1}{(z_{L+i}(E))^k} \Big(\frac{1}{t_i} \,+\, \mathcal{O}(E^{-2})\Big) P_i(z_{L+i}(E)) \\ 
= & \; \frac{t_i^{k-1}}{E^k} P^T_{i+L} \,+\, \mathcal{O}(E^{-k-1}) \;.
\end{align*}
Using the residue theorem one deduces from the above expression for the Riesz projections that
\begin{align*}
\mathcal{R}_I^E 
\;= & \; 
\begin{pmatrix} T & 0 \\ 0 & \one \end{pmatrix} 
\sum_{i \in I} \begin{pmatrix} \mathrm{Res}_{z_i(E)}((H(z) - E)^{-1}) & - \mathrm{Res}_{z_i(E)}(z^{-1}(H(z) - E)^{-1}) \\ 
\mathrm{Res}_{z_i(E)}(z^{-1}(H(z) - E)^{-1}) & - \mathrm{Res}_{z_i(E)}(z^{-2}(H(z) - E)^{-1}) \end{pmatrix} 
\begin{pmatrix} \one & 0 \\ 0 & R \end{pmatrix} \\ 
= & \; 
\begin{pmatrix} T & 0 \\ 0 & \one \end{pmatrix} 
\begin{pmatrix} \underset{ i \in I}{\sum} \frac{1}{t_{i-L}} P^T_{i} + \mathcal{O}(E^{-1}) & - \underset{ i \in I}{\sum} \frac{1}{E} P^T_{i} + \underset{i \in I}{\sum} \frac{1}{E} P^R_{i} + \mathcal{O}(E^{-2}) \\ 
\underset{i \in I}{\sum} \frac{1}{E} P^T_{i} - \underset{i \in I}{\sum} \frac{1}{E} P^R_{i} + \mathcal{O}(E^{-2}) & \underset{i \in I}{\sum} \frac{1}{r_i} P^R_{i} + \mathcal{O}(E^{-1}) \end{pmatrix} 
\begin{pmatrix} \one & 0 \\ 0 & R \end{pmatrix} \\ 
= & \; 
\begin{pmatrix} T & 0 \\ 0 & \one \end{pmatrix} 
\begin{pmatrix} T^{-1} P^T_I & \frac{1}{E} (P^R_I - P^T_I) \\ 
- \frac{1}{E} (P^R_I - P^T_I) & R^{-1} P^R_I \end{pmatrix} 
\begin{pmatrix} \one & 0 \\ 0 & R \end{pmatrix} + 
\begin{pmatrix} \mathcal{O}(E^{-1}) & \mathcal{O}(E^{-2}) \\ \mathcal{O}(E^{-2}) & \mathcal{O}(E^{-1}) \end{pmatrix} \\ 
= & \; 
\begin{pmatrix} P^T_I + \frac{1}{E} Q^T_I & \frac{1}{E} T (P^R_I - P^T_I) R \\ 
- \frac{1}{E} (P^R_I - P^T_I) & P^R_I + \frac{1}{E} Q^R_I \end{pmatrix} 
\,+\, \mathcal{O}(E^{-2}) 
\end{align*}
for some matrices $Q^T_I$ and $Q^R_I$, where the invertibility of the transfer matrix was used in order to choose a path $\gamma_I$ not encircling $0$. Since the Riesz projection is an idempotent, it follows that $Q^T_I$ and $Q^R_I$ are indeed off-diagonal w.r.t. $P^T_I$ and $P^R_I$ respectively.
\hfill $\Box$

\vspace{.2cm}

As a first application of Proposition~\ref{prop-RieszExpand1}, let us consider the $q$-functions used in \cite{KS}. They are defined in equation \eqref{eq-qTildeDef} below and will carry a tilde in order to distinguish them from $q_I$ and $\hat{q}_I$ (even though they can readily be connected to $\hat{q}_I$). The following proposition then gives their large $E$ asymptotics. For the special case $I=\{L+1, \dots, 2L\}$, the claim was already made in Lemma 36 of \cite{KS}. However, the argument provided there has a gap (more precisely, some residues were neglected), which is now filled by the proof given below.

\begin{proposition}
\label{prop-qtildeAsymp}
Let us consider
\begin{equation}
\label{eq-qTildeDef}
\tilde{q}_I(E) 
\;=\; 
{\det}_L \left( \begin{pmatrix} 0 \\ \one_L \end{pmatrix}^* \mathcal{R}_I^E \begin{pmatrix} \one_L \\ 0 \end{pmatrix} \right)
\;.
\end{equation}
It has the large $E$ asymptotics 
$$
\tilde{q}_I(E) 
\;=\; 
{\det}_L (P^T_I - P^R_I) E^{-L}\, +\, \mathcal{O}(E^{-L-1})
\;.
$$
\end{proposition}

\noindent {\bf Proof.} The claim follows directly from \eqref{eq-RieszExpansion} because $\tilde{q}_I$ extracts the lower left entry of  $\mathcal{R}_I^E$.
\hfill $\Box$

\vspace{.2cm}

Proposition~\ref{prop-qtildeAsymp} is only of interest if the cardinality of $I$ is equal to $L$. Indeed, if it is smaller, then \eqref{eq-RankSum} implies that  ${\det}_L (P^T_I - P^R_I)=0$, while if it is larger, the same holds by using the identity ${\det}_L (P^T_I - P^R_I)={\det}_L (P^T_{I^c} - P^R_{I^c})$. 

\vspace{.2cm}

In order to extract the leading coefficients of $\hat{q}_I$ and $q_I$, a preparatory lemma is needed which can be conveniently stated in terms of frames. Recall that a frame for a subspace  $\Ee\subset\CM^L$ of dimension $p$ is an injective matrix $\Phi\in\CM^{L\times p}$ with image $\Ran(\Phi) = \Ee$. Let us stress that the frame is {\it not} required to be orthonormalized, namely $\Phi^*\Phi=\one_p$ need not hold.  Now, given a (not necessarily selfadjoint) projection $P$ of rank $p$, one has a frame $\Phi\in\CM^{L\times p}$ for $\Ran(P)$ and another frame $\Phi^c\in\CM^{L\times (L-p)}$ for  $\Ran(\one-P)$.  Two further frames $\Psi\in\CM^{L\times p}$ and $\Psi^c\in\CM^{L\times (L-p)}$ are obtained by setting $(\Psi,\Psi^c) = \big((\Phi,\Phi^c)^{-1}\big)^*$. They satisfy
\begin{align*}
\Ran(\Psi)
&
\;=\;
\Ker (\Psi^*)^\perp
\;=\;
\Ran(\Phi^c)^\perp
\;=\;
\Ran(\one-P)^\perp
\;=\;
\Ran(P^*)
\;,
\end{align*}
where the second equality follows, because $\Psi^* \Phi^c = 0$ and $\rk(\Phi^c) = \dim (\Ker (\Psi^*))$. In the same manner one gets $\Ran(\Psi^c) = \Ran(P)^\perp = \Ran(\one - P^*)$. Now the projection can be written as $P=\Phi \Psi^*$ ({\it e.g.} Section~5.1  in \cite{DSW} combined with the facts that here $\Ran(P^*)=\Ker(P)^\perp$ and $\Psi^*\Phi=\one_p$) and similarly $\one-P=\Phi^c (\Psi^c)^*$. Moreover, one has
$$
(\Psi,\Psi^c)^*\, P\,(\Phi,\Phi^c)
\;=\;
(\Phi,\Phi^c)^{-1}\, P\,(\Phi,\Phi^c)
\;=\;
\begin{pmatrix}
\one_p & 0 \\ 0 & 0
\end{pmatrix}
\;.
$$

\begin{lemma}
\label{lem-qAsymp}
Let $P\in\CM^{L\times L}$ be a projection of rank $p=\rk(P)$ with associated frames $\Phi$ and $\Phi^c$ for $\Ran(P)$ and $\Ran(\one-P)$ as well as frames $\Psi\in\CM^{L\times p}$ and $\Psi^c\in\CM^{L\times (L-p)}$ defined by $(\Psi,\Psi^c) = \big((\Phi,\Phi^c)^{-1}\big)^*$. Then for any matrix valued function $E\in\CM\mapsto M^E\in\CM^{L\times L}$ satisfying at infinity $M^E = M_0 + \Oo(E^{-1})$, one has 
$$
{\det}_L(E\,P+M^E)
\;=\;
{\det}_{L-p}\big((\Psi^c)^* M_0 \Phi^c \big) E^p \big(1+\Oo(E^{-1})\big)
\;.
$$
\end{lemma}

\noindent {\bf Proof.} Due to the above identities,
\begin{align*}
& 
{\det}_L(E\,P+M^E)
\;=\;
{\det}_L
\big((\Psi,\Psi^c)^*(E \, P + M^E)(\Phi,\Phi^c)\big)
\\
&
\;=\;
{\det}_L
\Big(
\begin{pmatrix}
E\,\one_p + \Psi^* M^E \Phi & \Psi^* M^E \Phi^c
\\
(\Psi^c)^* M^E \Phi & (\Psi^c)^* M^E \Phi^c
\end{pmatrix}
\Big)
\\
&
\;=\;
{\det}_p
\big(
E\,\one_p+\Phi^* M^E \Phi
\big) 
{\det}_{L-p}
\big(
(\Psi^c)^* M^E \Phi^c
-
(\Psi^c)^* M^E \Phi
(E\,\one_p + \Psi^* M^E \Phi)^{-1}
\Psi^* M^E \Phi^c
\big)
\\
&
\;=\;
E^p
\,{\det}_{L-p}
\big(
(\Psi^c)^* M_0 \Phi^c
\big)
\big(
1+\Oo(E^{-1})
\big)
\;,
\end{align*}
showing the claim.
\hfill $\Box$

\vspace{.2cm}

Next let us address the functions $\hat{q}_I$ used in Theorem~\ref{theo-boundary}. For similar reasons as to why the leading coefficient of the expansion of the $\tilde{q}_I$ are zero for $|I| \neq L$, the leading coefficient in the following expansion can be non-vanishing only if $|I|\geq L$.

\begin{proposition}
\label{prop-qhatAsymp}
Set $p=\rk(P^T_I)$ and let $\Phi\in\CM^{L\times p}$ and $\Phi^c\in\CM^{L\times (L-p)}$ be frames for $\Ran(P^T_{I})$ and $\Ran(P^T_{I^c})$ respectively and $\Psi^c\in\CM^{L\times (L-p)}$ be the associated frame for $\Ran((P^T_{I^c})^*)$. Then the function $\hat{q}_I$ defined in \eqref{eq-qDef} has the large $E$ asymptotics 
$$
\hat{q}_I(E) 
\;=\; 
{\det}_{L-p}\big(
(\Psi^c)^* P^R_I (C-V) \Phi^c \big) E^{-L+p} \big(1+\Oo(E^{-1})\big)
\;.
$$
\end{proposition}

\noindent {\bf Proof.} Due to Proposition~\ref{prop-RieszExpand1} there are matrices $a$, $b$, $c$ and $d$, whose entries are bounded functions of $E$ for all large $E$, such that 
\begin{equation}
\label{eq-ExpansionRieszProj}
\mathcal{R}_I^E 
\;=\; 
\begin{pmatrix} P^T_I + \frac{1}{E} Q^T_I + \frac{1}{E^2} a & \frac{1}{E} T (P^R_I - P^T_I ) R + \frac{1}{E^2} b \\ 
- \frac{1}{E} (P^R_I - P^T_I ) + \frac{1}{E^2} c & P^R_I + \frac{1}{E} Q^R_I + \frac{1}{E^2} d \end{pmatrix} 
\;.
\end{equation}
Notice that $\mathcal{R}_I^E$ and $\Tt^E$ commute. By computing the commutator of their expansions in $E^{-1}$, one finds numerous identities. In particular, for the term of order $-1$ in the lower left corner one finds 
$$
T^{-1} Q^T_I T 
\;=\;
c + (P^R_I - P^T_I) V + Q^R_I 
\;.
$$
Using this identity one obtains
\begin{align*}
\hat{q}_{I}(E)
&
\;=\;
{\det}_L\Big(
\binom{0}{\one}^*
{\Rr}_{I}^E
\begin{pmatrix}
E\,\one - C & - \one \\ \one & 0 
\end{pmatrix}
\binom{\one}{0}\Big)
\\
&
\;=\;
{\det}_L \Big( E P^T_I + (P^R_I - P^T_I )C + Q^R_I  + c + \frac{1}{E} (- c C + d) \Big) E^{-L} 
\\
&
\;=\;
{\det}_L \Big( E P^T_I + (P^R_I - P^T_I ) (C - V) + T^{-1} Q^T_I T + \frac{1}{E} (- c C + d) \Big) E^{-L} 
\;.
\end{align*}
An application of Lemma~\ref{lem-qAsymp} now gives the desired result, {\it i.e.} there exists a frame $\Psi^c\in\CM^{L\times (L-p)}$ for $\Ran((P^T_{I^c})^*)$ such that
\begin{align*}
\hat{q}_{I}(E)
&
\;=\;
{\det}_{L-p} \Big( (\Psi^c)^* \left( (P^R_I - P^T_I ) (C - V) + T^{-1} Q^T_I T \right) \Phi^c \Big) E^{-L+p} \big(1+\Oo(E^{-1})\big) 
\\
&
\;=\;
{\det}_{L-p} \Big( (\Psi^c)^* P^R_I (C - V) \Phi^c \Big) E^{-L+p} \big(1+\Oo(E^{-1})\big) 
\;,
\end{align*}
since $\Phi^c = P^T_{I^c} \Phi^c$, $(\Psi^c)^* = (\Psi^c)^* P^T_{I^c}$, $P^T_{I^c} P^T_I =0$ and $ P^T_{I^c} Q^T_I P^T_{I^c} = 0$ by Proposition~\ref{prop-RieszExpand1}.
\hfill $\Box$

\begin{remark}
{\rm
For $C=V$, Proposition~\ref{prop-qhatAsymp} is void. In this special case, one has
\begin{align*}
\hat{q}_I(E)
&
\;=\;
{\det}_L(T)\;
{\det}_L\Big(
\binom{0}{\one}^*
{\Rr}_{I}^E
\begin{pmatrix}
((E\,\one-V)T^{-1} & - R \\ T^{-1} & 0 
\end{pmatrix}
\binom{\one}{0}\Big)
\\
&
\;=\;
{\det}_L(T)\;
{\det}_L\Big(
\binom{0}{\one}^*
\begin{pmatrix}
((E\,\one-V)T^{-1} & - R \\ T^{-1} & 0 
\end{pmatrix}
{\Rr}_{I}^E
\binom{\one}{0}\Big)
\\
&
\;=\;
{\det}_L\Big(
\binom{\one}{0}^*
{\Rr}_{I}^E
\binom{\one}{0}\Big)
\;.
\end{align*}
It is not possible to extract the leading order coefficient directly from \eqref{eq-RieszExpansion}, but one {can expand} $\Tt^E$ in its Riesz projections and then apply Lemma~32 from \cite{KS} to compute the highest order term. This leads to a supplementary factor $Z_I(E)$ in \eqref{eq-OpenIdentityIntro}, in agreement with the formulas in the proof of Theorem~31 in \cite{KS}.
}
\hfill $\diamond$
\end{remark}

Finally let us provide the leading order coefficient for the functions $q_I$ defined in \eqref{eq-qDef} and appearing in Theorem~\ref{thm-LimitSpecIntro}. 

\begin{proposition}
\label{prop-qAsymp}
Set $p=\rk(P^T_I)$ and $\hat{p}=\rk(P^R_I)$. Let $\Phi,\Psi\in \CM^{L\times p}$ and $\Phi^c,\Psi^c\in\CM^{L\times (L-p)}$ be the frames associated to $P^T_{I}$, and similarly $\hat{\Phi},\hat{\Psi}\in \CM^{L\times \hat{p}}$ and $\hat{\Phi}^c,\hat{\Psi}^c\in\CM^{L\times (L-\hat{p})}$ those associated to $P^R_{I}$. Then
\begin{align*}
q_I(E) 
\;=\;
\frac{{\det}_{L-p}\big((\Psi^c)^* B \Phi^c \big)
\,{\det}_{\hat{p}}
\big(\hat{\Psi}^* A \hat{\Phi}
\big)}
{(-1)^{p-\hat{p}} {\det}_L(B)}
\,
E^{p-\hat{p}} \big(1+\Oo(E^{-1})\big)
\;.
\end{align*}
\end{proposition}

\noindent {\bf Proof.} Let us set
$$
\Mm^E 
\;=\; 
\big(\mathcal{R}_I^E ( \mathcal{T}_{\mbox{\rm\tiny bd}}^E + \one_{2L}) - \one_{2L} \big) \,\mathrm{diag}(B, \one_{L}) 
\;,
$$
as then
$$
q_I (E) 
\;=\; 
\det\!_{2L}(\mathcal{R}_I^E \mathcal{T}_{\mbox{\rm\tiny bd}}^E - \mathcal{R}_{I^c}^E) 
\;=\; \det\!_{2L}(\mathcal{R}_I^E ( \mathcal{T}_{\mbox{\rm\tiny bd}}^E + \one_{2L} ) - \one_{2L})
\;=\; 
{\det}_L(B)^{-1} \det\!_{2L}(\Mm^E) \;.
$$
The determinant of $\Mm^E$ will be computed by applying Lemma~\ref{lem-qAsymp} twice. Using again equation \eqref{eq-ExpansionRieszProj} for $\mathcal{R}_I^E$ one has
\begin{align*}
\Mm^E 
\;= & \;
\begin{pmatrix} P^T_I + \frac{1}{E} Q^T_I + \frac{a}{E^2} & \frac{1}{E} T (P^R_I - P^T_I) R + \frac{b}{E^2} \\ 
- \frac{1}{E} (P^R_I - P^T_I) + \frac{c}{E^2} & P^R_I + \frac{1}{E} Q^R_I + \frac{d}{E^2} \end{pmatrix}
\begin{pmatrix}E + B - C & - A \\ 
\one_L & \one_L \end{pmatrix} - 
\begin{pmatrix} B & 0 \\ 
0 & \one_L \end{pmatrix} \\
= & \; 
E \begin{pmatrix} P^T_I & 0 \\ 0 & 0 \end{pmatrix} + 
\begin{pmatrix} P^T_I (B - C) + Q^T_I - B & - P^T_I A \\
P^T_I & - P^R_{I^c} \end{pmatrix} \\
& \;+ \,
\frac{1}{E} \begin{pmatrix} Q^T_I (B - C) + a + T (P^R_I - P^T_I) R & - Q^T_I A + T (P^R_I - P^T_I) R \\ 
- (P^R_I - P^T_I)(B - C) + c + Q^R_I & (P^R_I - P^T_I) A + Q^R_I \end{pmatrix}\, +\,
\mathcal{O}(E^{-2}) \;. 
\end{align*}
To shorten notations, let us write $\Mm^E=E\Mm_1+\Mm_0+E^{-1}\Mm_{-1}+\Oo(E^{-2})$ for coefficient matrices $\Mm_j$ which can be read off from the last equation. Note that $\Mm_1$ is a projection of rank $p=\rk(P^T_I)$. Then $\tilde{\Phi}^c=\binom{\Phi^c\;0}{0\;\;\one_L}\in\CM^{2L\times (2L-p)}$ is a frame for $\Ran(\one_{2L} - \Mm_1)$, and $\tilde{\Psi}^c = \binom{\Psi^c\;0}{0\;\;\one_L}$ is a frame for $\Ran(\one_{2L} - \Mm^*_1)$. By Lemma~\ref{lem-qAsymp} one then has
\begin{align*}
q_I (E)\, {\det}_L(B)
&
\;=\; 
{\det}_{2L}\big(E \, \Mm_1 + \Mm_0 + E^{-1}\Mm_{-1} + \Oo(E^{-2}) \big)
\\
&
\;=\;
{\det}_{2L-p}
\Big(
(\tilde{\Psi^c})^* \Mm_0 \tilde{\Phi}^c
\Big)
E^p\,
\big(1+\Oo(E^{-1})\big)
\;.
\end{align*}
Now one can compute the first summand in the determinant, using $\Phi^c = P^T_{I^c} \Phi^c$, $(\Psi^c)^* = (\Psi^c)^* P^T_{I^c}$,   $P^T_{I^c} P^T_{I} = P^T_{I} P^T_{I^c} = $ and $P^T_{I^c} Q^T_I P^T_{I^c} = 0$:
\begin{align*}
(\tilde{\Psi}^c)^* \Mm_0 \tilde{\Phi}^c
&
\;=\;
\begin{pmatrix} - (\Psi^c)^* B \Phi^c & 0 \\
0 & - P^R_{I^c} \end{pmatrix}
\;.
\end{align*}
The determinant of this matrix vanishes when $I \cap \{1, \dots , L\}$ is non-empty, because then ${\det}_L(P^R_{I^c})=0$. To get the leading order coefficient for $q_I$, one hence needs to also take lower order contributions into account. There are two sources for such terms, namely $\Mm_{-1}$ and the lower order term in the Schur complement used in the proof of Lemma~\ref{lem-qAsymp}. Supposing that $(\Psi^c)^* B \Phi^c$ is invertible, let us set
\begin{align*}
\hat{\Mm}_{-1}
\;=\;
\begin{pmatrix} - (\Psi^c)^* B \Phi^c & 0 \\
0 & -\one_L \end{pmatrix}^{-1}
(\tilde{\Psi}^c)^* \left(
\Mm_{-1} 
- 
\Mm_{0} \tilde{\Phi} 
\left( (\tilde{\Psi})^* \Mm_{1} \tilde{\Phi} \right)^{-1} 
(\tilde{\Psi})^* \Mm_{0}
\right) \tilde{\Phi}^c
\;,
\end{align*}
so $\hat{\Mm}_{-1} \in \CM^{(2L-p)\times(2L-p)}$.  Using that $\tilde{\Phi}=\binom{\Phi}{0}$ and $\tilde{\Psi}=\binom{\Psi}{0}$, one finds that $(\tilde{\Psi})^* \Mm_{1} \tilde{\Phi}=\Psi^*\Phi=\one_p$ so that
\begin{align*}
\hat{\Mm}_{-1}
\;=\;
\begin{pmatrix} - (\Psi^c)^* B \Phi^c & 0 \\
0 & -\one_L \end{pmatrix}^{-1}
(\tilde{\Psi}^c)^* \left(
\Mm_{-1} 
- 
\Mm_{0} 
\begin{pmatrix}
P^T_I & 0 \\ 0 & 0
\end{pmatrix}
\Mm_{0}
\right) \tilde{\Phi}^c
\;.
\end{align*}
Replacing  one then has 
\begin{align*}
q_I (E) {\det}_L(B)
&
\;=\;
(-1)^{p}
\,
{\det}_{L-p}\big((\Psi^c)^* B \Phi^c \big)
\\
&
\;\;\;\;\;\;\;\cdot {\det}_{2L-p}
\Big( E 
\begin{pmatrix} \one_{L-p} & 0 \\
0 & P^R_{I^c} \end{pmatrix}
+
\hat{\Mm}_{-1}
+
\Oo(E^{-1})
\Big) 
E^{-2L+2p} 
\big(1+\Oo(E^{-1})\big)
\;.
\end{align*}
Next the second determinant can be computed using again Lemma~\ref{lem-qAsymp}, now for the projection $P=\diag(\one_{L-p}, P^R_{I^c})\in\CM^{(2L-p)\times (2L-p)}$. The associated frames for $P$ are denoted by $\check{\Phi},\check{\Psi}\in\CM^{(2L-p)\times (2L-p-\hat{p})}$ and $\check{\Phi}^c,\check{\Psi}^c\in\CM^{(2L-p)\times \hat{p}}$. They are explicitly given by
$$
\check{\Phi}
\;=\;
\begin{pmatrix}
\one_{L-p} & 0 \\ 0 & \hat{\Phi}^c
\end{pmatrix}
\;,
\qquad
\check{\Phi}^c
\;=\;
\begin{pmatrix}
0 \\ \hat{\Phi}
\end{pmatrix}
\;,
\qquad
\check{\Psi}
\;=\;
\begin{pmatrix}
\one_{L-p} & 0 \\ 0 & \hat{\Psi}^c
\end{pmatrix}
\;,
\qquad
\check{\Psi}^c
\;=\;
\begin{pmatrix}
0 \\ \hat{\Psi}
\end{pmatrix}
\;.
$$
Hence Lemma~\ref{lem-qAsymp} implies
\begin{align*}
{\det}_{2L-p}
\Big( E 
\begin{pmatrix} \one_{L-p} & 0 \\
0 & P^R_{I^c} \end{pmatrix}
+
\hat{\Mm}_{-1}
+
\Oo(E^{-1})
\Big)
&
\;=\;
{\det}_{\hat{p}}\big( (\check{\Psi}^c)^* \hat{\Mm}_{-1} \check{\Phi}^c \big) 
E^{2L-p-\hat{p}}
\big(1+\Oo(E^{-1})\big)
\;.
\end{align*}
Then one can carefully compute
\begin{align*}
q_I (E)\, {\det}_L(B)
&
\;=\; 
(-1)^{p}
\,
{\det}_{L-p}\big((\Psi^c)^* B \Phi^c \big) 
\,
{\det}_{\hat{p}}\big( 
(\check{\Psi}^c)^* \hat{\Mm}_{-1} \check{\Phi}^c  \big)
\,
E^{p-\hat{p}}
\,
\big(1+\Oo(E^{-1})\big)
\\
&
\;=\;
(-1)^{p}
\,
{\det}_{L-p}\big((\Psi^c)^* B \Phi^c \big) 
\\
&
\;\;\;\;\;\;\;\; \cdot 
{\det}_{\hat{p}}\big( - \hat{\Psi}^* ((P^R_I - P^T_I) A + Q^R_I + \Phi \Psi^* A) \hat{\Phi} \big)
\,
E^{p-\hat{p}}
\,
\big(1+\Oo(E^{-1})\big)
\\
&
\;=\;
(-1)^{p-\hat{p}}
\,
{\det}_{L-p}\big((\Psi^c)^* B \Phi^c \big) 
\,
{\det}_{\hat{p}}\big(\hat{\Psi}^* A \hat{\Phi} \big)
\,
E^{p-\hat{p}}
\,
\big(1+\Oo(E^{-1})\big)
\;,
\end{align*}
since $\hat{\Psi}^* = \hat{\Psi}^* P^R_I$, $\Psi = P^R_I \Psi$, $P^R_I Q^R_I P^R_I = 0$ and $\Phi \Psi^* = P^T_I$. This is precisely the claim. The case in which $(\Psi^c)^* B \Phi^c$ is singular, is covered by a  perturbation argument. 
\hfill $\Box$

\vspace{.2cm}

\noindent
{\bf Proof} of Theorem~\ref{theo-Expansion}:  Let us focus on the statement about $E\mapsto q_I(E)$, because the other case is treated similarly. First note that for Lebesgue almost all $R$ and $T$, $R$ and $T$ are invertible and have simple spectrum. In that case it follows from Lemma 24 in \cite{KS} that $\mathcal{F}$ is finite. One can thus safely assume that $\mathcal{F}$ is finite. Recall that
$$
q_I (E) 
\;=\; 
\det\!_{2L}\big(\mathcal{R}_I^E \mathcal{T}_{\mbox{\rm\tiny bd}}^E - \mathcal{R}_{I^c}^E\big) 
\;=\; \det\!_{2L}\big(\mathcal{R}_I^E ( \mathcal{T}_{\mbox{\rm\tiny bd}}^E + \one_{2L} ) - \one_{2L}\big) 
\;.
$$
One can now use the analyticity properties of the Riesz projections $\mathcal{R}_I^E$ to obtain the desired result in the following way.  Note that $\mathcal{T}_{\mbox{\rm\tiny bd}}^E$ is an entire function in $E$ and the Riesz projections are locally holomorphic, meaning holomorphic up to a choice of branch cut in $E$ on $\CM \setminus \mathcal{A}$, see \eqref{eq-AmbiguitySet} for the definition of $\mathcal{A}$. On the set $\mathcal{A}$ the functions $\mathcal{R}_I^E$ can exhibit discontinuities due to the relabeling that occurs. However, if one disregards this relabeling and appropriately glues the functions $\mathcal{R}_I^E$ together, they are actually locally holomorphic in $E$ on $\CM \setminus \mathcal{F}$. The same thus also holds for the $q_I$, {\it i.e.} the functions $q_I$ are holomorphic on $\CM \setminus \mathcal{F}$ up to relabeling and a choice of branch cut. Moreover, since the $\mathcal{R}_I^E$ are analytic at infinity, the branch cut can be chosen to be bounded. 

\vspace{.1cm}

Note that the leading coefficient in the expansion of $q_I$ as computed in Proposition~\ref{prop-qAsymp} is non-zero for Lebesgue almost all choices of matrices $R$, $T$, $A$ and $B$. Moreover this is still the case if one adds the condition that $R$ and $T$ need to be invertible and have simple spectrum, since all these conditions can be expressed as certain polynomials in the entries of the matrices being non-zero. It follows that the functions $q_I$, analytically continued through the set $\mathcal{A}$, are not-identically-zero holomorphic functions on the complement of the branch cut $\mathcal{V}$. Their zero sets will thus be discrete. Since the analytic continuation through $\mathcal{A}$ is done by relabeling, all the locally holomorphic functions $q_I$ themselves will have discrete zero sets on $\CM \setminus \mathcal{V}$. By choosing different branchs cuts $\mathcal{V}$, one sees that the locally holomorphic $q_I$ actually will have discrete zero sets on $\CM \setminus \mathcal{F}$. 
\hfill $\Box$

\appendix

\section{Density of diagonalizable Toeplitz matrices}
\label{sec-Density}

This appendix proves that within the set of Toeplitz matrices of the form \eqref{eq-ToeplitzIntro} the set of diagonalizable ones is dense w.r.t. the Euclidean topology on the set of matrix entries. This fact is used in the derivation of Widom's formula. For sake of concreteness, we will focus on the set $\Toep_{L,N}$ of the matrices $H_N(R,T,V)$ given by  \eqref{eq-ToeplitzIntro} with $R,V,T \in \CM^{L\times L}$ and $R$ and $T$ invertible. Moreover, let $\DToep_{L,N}$ denote the set of diagonalizable elements in $\Toep_{L,N}$. The case where $(A,B,C)\not=(R,T,V)$, but fixed, is dealt with in a similar manner and leads to the same result. 

\begin{proposition} 
\label{prop-density}
The set $\DToep_{L,N}$ is dense in $\Toep_{L,N}$ w.r.t. the subspace topology induced by the Euclidean topology on $\mathbb{C}^{NL \times NL}$.
\end{proposition}

The proof of this result is inspired by the standard algebrogeometric argument for the fact that the set of diagonalizable matrices is dense in the set of all square matrices (of fixed size over {an algebraically} closed infinite field). This will use a number of well-known facts from algebraic geometry which can be found, for example, in \cite{Mum} or \cite{Kra}.

\vspace{.2cm}

\noindent {\bf Proof.}
Consider the set $\mathrm{Circ}_{L,N} \subset \CM^{NL \times NL}$ of block circulant matrices associated to symbols of the form $z \mapsto z^{-1}R+ V + z T$ with $R,V,T \in \CM^{L\times L}$. Note that $\Toep_{L,N} \subset \mathrm{Circ}_{L,N}$ with $R$ and $T$ being invertible. One can exhibit $\mathrm{Circ}_{L,N}$ as an algebraic set in the following way:
$$
\mathrm{Circ}_{L,N} 
\,=\; 
\{ M \in \CM^{NL \times NL}\,:\, f(M) = 0 \text{ for all } f \in \langle S_{\mbox{\rm\tiny Circ}} \rangle \} \,,
$$
where $S_{\mbox{\rm\tiny Circ}} \subset \mathbb{C}[x_{1,1}, \dots, x_{NL,NL}]$ is the set of polynomials
$$ 
S_{\mbox{\rm\tiny Circ}} 
=\, 
\{x_{i,j} - x_{nL+i \,\mathrm{mod}\, NL, nL+j \,\mathrm{mod}\, NL} \colon \left\lceil \tfrac{i}{L} \right\rceil + \left\lceil \tfrac{j}{L} \right\rceil \leq 3 \text{ and all } n \} 
\cup 
\{ x_{i,j} \colon 1 < \left| \left\lceil \tfrac{i}{L} \right\rceil - \left\lceil \tfrac{j}{L} \right\rceil \right| < N  \} 
\,.
$$
Note that the coordinate ring of $\mathrm{Circ}_{L,N}$ is given by
$$
\mathbb{C}[x_{1,1}, \dots, x_{NL,NL}] / \langle S_{\mbox{\rm\tiny Circ}} \rangle 
\;\cong\; 
\mathbb{C}[x_1, \dots, x_{3L^2}] \;.
$$
Since this is an integral domain, it follows that $\mathrm{Circ}_{L,N}$ is an affine algebraic variety, {\it i.e.} an irreducible algebraic set. Next let us consider the maps ${\det}_R, {\det}_T \colon \mathrm{Circ}_{L,N} \to \mathbb{C}$ given by ${\det}_T (M) = \det(M_{i,L+j})_{i,j=1}^L$ and ${\det}_R (M) = \det(M_{L+i,j})_{i,j=1}^L$ for $M = (M_{i,j})_{i,j=1}^{NL} \in \mathrm{Circ}_{L,N}$. Then
$$
\Toep_{L,N} 
\;=\; 
{\det}_R^{-1}(\mathbb{C}\setminus\{0\}) \cap {\det}_T^{-1}(\mathbb{C}\setminus\{0\}) \;. 
$$
Further let us consider the map $d \colon \mathrm{Circ}_{L,N} \to \mathbb{C}$ that sends a matrix to the discriminant of its characteristic polynomial. Recall that a matrix has simple spectrum if and only if the discriminant of its characteristic polynomial is non-zero. Therefore it follows that
$$
d^{-1}(\mathbb{C}\setminus \{0\}) \cap \Toep_{L,N} \;\subset\; \DToep_{L,N} \;. 
$$
In what follows, it will be shown that $d^{-1}(\mathbb{C}\setminus \{0\}) \cap \Toep_{L,N}$ is non-empty and open w.r.t. the Zariski topology on $\mathrm{Circ}_{L,N}$. It then follows from general algebrogeometric arguments that it is dense in $\mathrm{Circ}_{L,N}$ w.r.t. the subspace topology inherited from the Euclidean topology on $\mathbb{C}^{NL \times NL}$ ({\it e.g.} Theorem 1 in Section I.10 of \cite{Mum}, or Appendix C in \cite{Kra}). Therefore one can conclude that $\DToep_{L,N}$ is dense in $\Toep_{L,N}$ w.r.t. its subspace topology.

\vspace{.1cm}

Note that all three maps ${\det}_R$, ${\det}_T$ and $d$ are given by polynomial expressions in the entries of the matrix argument. It follows that these maps are continuous w.r.t. the Zariski topology. Since $\mathbb{C}\setminus \{0\}$ is Zariski open, it follows that 
$$
d^{-1}(\mathbb{C}\setminus \{0\}) \cap \Toep_{L,N} 
\;=\;  
d^{-1}(\mathbb{C}\setminus \{0\}) \cap {\det}_R^{-1}(\mathbb{C}\setminus\{0\}) \cap {\det}_T^{-1}(\mathbb{C}\setminus\{0\}) 
$$
is also Zariski open. To show that this set is non-empty, consider $H_N=H_N(R,T,V)$ with block entries $R = \one_L$, $V = \mathrm{diag}(7, 14, 21, \dots, 7L)$ and $T = 2 \one_L$. Clearly $H_N \in \mathrm{Circ}_{L,N}$. Note that ${\det}_R(H_N) = \det(R)=1$ and ${\det}_T(H_N) = \det(T)=2^L$, so $H_N \in \Toep_{L,N}$. Moreover, $H_N$ decomposes into a direct sum $\bigoplus_{j=1}^L h_j$, where the $h_j$ are $N\times N$ circulant matrices with symbols $h_j(z)= z^{-1} + 7j + 2z$. The spectrum of $h_j$ is then given by 
%
$$
\spec(h_j) 
\;=\; 
\big\{ e^{- \frac{2 \pi \imath k}{N}} + 7j + 2 e^{\frac{2 \pi \imath k}{N}} \colon 1 \leq k \leq N 
\big\} 
\;. 
$$
It follows that $H_N$ has simple spectrum, so $d(H_N) \neq 0$. Therefore $d^{-1}(\mathbb{C}\setminus \{0\}) \cap \Toep_{L,N}$ is indeed non-empty.
\hfill $\Box$

\vspace{.2cm}

 \noindent {\bf Acknowledgements:} This work was supported by the DFG grant SCHU 1358/8-1. Data sharing not applicable to this article as no datasets were generated or analyzed during the current study. The authors have no competing interests to declare that are relevant to the content of this article.


\end{document}